\documentclass{emulateapj}
\usepackage{graphicx}

\newcommand{\kms}{${\rm km \; s^{-1}}$}
\newcommand\keff{\kappa_{\rm eff}}
\newcommand\geff{\gamma_{\rm eff}}

\shorttitle{Lens Galaxy Environments}
\shortauthors{Momcheva et al.}

\begin{document}
\singlespace

\title{A Spectroscopic Study of the Environments of Gravitational Lens Galaxies}
\author{Ivelina Momcheva}
\author{Kurtis Williams}
\affil{Steward Observatory, 933 N. Cherry Ave., Tucson, AZ 85721}
\author{Charles Keeton}
\affil{Department of Physics and Astronomy, Rutgers University,
  Piscataway, NJ 08854}
\and
\author{Ann Zabludoff}
\affil{Steward Obs., 933 N. Cherry Ave., Tucson, AZ 85721}

\begin{abstract}

We present the first results from our spectroscopic survey\footnote{This paper includes data gathered with the 6.5 meter Magellan Telescopes located at Las Campanas Observatory, Chile.} of the
environments of strong gravitational lenses.  The lens galaxy belongs
to a poor group of galaxies in six of the eight systems in our sample.
We discover three new groups associated with the lens galaxies of
BRI 0952$-$0115 (five members), MG 1654+1346 (seven members), and
B2114+022 (five members).  We more than double the number of members
for another three previously known groups around the lenses
MG 0751+2716 (13 total members), PG 1115+080 (13 total members), and
B1422+231 (16 total members).  We determine the kinematics of the six
groups, including their mean velocities, velocity dispersions, and
projected spatial centroids.  For the newly discovered groups, we
quantify these properties for the first time.  For the other three
groups, the increased membership allows us to make more robust
estimates of the kinematic properties of the groups than previously
possible.  The velocity dispersions of the groups range from
$110^{+170}_{-80}$ to $470^{+100}_{-90}$ \kms. The higher velocity
dispersions (for the richer groups MG0751, PG1115, and B1422) are
consistent with those of nearby X-ray luminous groups, while the
others (for the poorer groups BRI0952, MG1654, and B2114) are more
typical of nearby dynamically younger groups.  The lens galaxy is
the brightest member in fewer than half of the groups.  In general,
the brightest group galaxy is an early-type galaxy that lies off
the center of the potential and occupies an orbit indistinguishable
from the other group members.  In at least three of the lenses ---
MG0751, PG1115, and B1422 --- the group environment significantly
affects the lens potential.  These lenses happen to be the
quadruply-imaged ones in our sample, which suggests a connection
between image configuration and environment. Finally, our survey
allows us to assess for the first time whether mass structures along
the line of sight are important for lensing.  We first show that,
in principle, the lens potential may be affected by line-of-sight
structures over a wide range of spatial and redshift offsets from
the lens.  We then quantify real line-of-sight effects using our
survey and find that at least four of the eight lens fields have
substantial interloping structures close in projection to the
lens, and at least one of those structures (in the field of MG0751)
significantly affects the lens potential.

\end{abstract}

\keywords{gravitational lensing --- (galaxies:) quasars: individual
(MG 0751+2716, BRI 0952$-$0115,  PG 1115+080, B1422+231, MG 1654+1346,
PMN J2004$-$1349, B2114+022, HE 2149$-$2745) --- galaxies: clusters:
general --- galaxies: halos}

\section{Introduction}

The study of strong gravitational lens systems offers critical
constraints on the masses, shapes, evolution, and substructure of
galaxy dark matter halos \citep[e.g.,][]{csk91,keetonoptical-98,MM,
DK,rusin-03,rusin-05,TK04,ferreras}, on the Hubble constant
independent of the local distance ladder \citep[e.g.,][]{refsdal,
kochanek-schechter}, and on the dark energy density \citep[e.g.,][]
{turner-90,csk96a,chae-stats,linder,mitchell}.  However, our
understanding of observed lenses is limited by uncertainties and
biases in the lens models necessary to analyze the data.  Despite
improving data for lensed images and lens galaxies, astrophysical
applications of lensing are still hindered by poor knowledge of
the environments in which strong lens systems reside.

Several arguments suggest that lenses have complex environments.
Statistical arguments based on galaxy demographics imply that at
least 25\% of lens galaxies lie in dense environments such as groups
and clusters \citep{kcz}.  From spectroscopic observations, several
lenses are in fact known to lie in groups \citep[MG 0751+2716,
PG 1115+080, B1422+231, and B1608+656;][]{T&Kmg0751-99,kundicpg1115,
kundicb1422-97,tonrypg1115,flb1608}, and several others in clusters
\citep[RX J0911+0551, Q0957+561, HST 14113+5221, and MG 2016+112;][]
{rxj0911,young0957,hst1411,mg2016+112}.  Indirect evidence for the
existence of other groups comes from the large tidal shears required
to explain the image configurations of many four-image (quad) lenses,
which presumably come from mass structures near the lens galaxy or
along the line of sight \citep{keeton-97}.  The range of required
shears in quad lenses could reflect a range of environment densities,
running from poor groups to rich clusters.  Comparisons of the lensing
rate in different surveys have also been cited as evidence that many
lens galaxies probably lie in groups \citep{blandford}.  Finally,
theoretical models predict that lens galaxies reside in complex
environments that produce substantial shears, although it is not yet
clear whether the models predict shears large enough to explain real
quad lenses \citep[][]{holder,dalal}.

If not handled properly, complex environments can inject
uncertainties and biases into the astrophysical quantities
derived from lens models \citep[see][hereafter KZ04]{kz04}.  For
example, neglecting environment altogether leads to lens models
that, for most purposes,\footnote{The important exception is
measurements of the total mass within the Einstein radius, which
are largely independent of assumptions built into lens models
\citep[e.g.,][]{csk91,cohn1933}.} are simply wrong.  Approximating
environmental effects with a simple shear term leads to models
that are better but still tend to overestimate the Hubble constant,
the velocity dispersion of the lens galaxy, and the dark energy
density $\Omega_\Lambda$, and to underestimate the magnifications
of the lensed images.  In principle, modeling the full
richness of environmental effects can remove these biases, and may
also resolve the long-standing puzzle of why quad lenses are
almost as common as doubles in statistically complete lens samples
\citep[see][KZ04]{kingQD,csk96b,keeton-97,rusin,cohn}.  Such an
analysis requires detailed knowledge of the galaxy populations,
velocity dispersions, and projected spatial centroids of groups
and clusters around lenses in order to determine how the environments
affect the lens potentials.  To date, such observations have mainly
been carried out for the few lenses that reside in clusters, which
leaves many lenses whose environments are known poorly or not at
all.  Worse, existing observations cannot characterize the
{\em distribution} of lens environments, so we cannot assess
environment-related biases in statistical quantities (such as
$\Omega_\Lambda$ or the quad/double ratio) or ensemble properties
(such as evolution or substructure).  While the environment
distribution can be predicted from theoretical models
\citep[][]{keeton-97,holder,dalal}, disagreements among the
models, and discrepancies between the predicted distributions
and the shears required to fit observed lenses, raise questions
about the predictions.

These issues have not been adequately addressed with observations,
because no systematic survey of lens environments exists.  Surveys of
a few lens fields have been published individually or in pairs 
\citep{young0957,kundicpg1115,kundicb1422-97,hst1411,tonrypg1115,
T&Kmg0751-99,rxj0911,fl,flb1608,mg2016+112}.  In many cases, though,
those surveys only spanned a $\sim\!30\arcsec$ field around each
lens, so they did not adequately sample group or cluster membership
out to the virial radius ($\sim\!0.7$ Mpc for groups, corresponding
to $\sim\!3\arcmin$ at the redshifts of the lenses we study).  We
have undertaken a systematic deep and wide-field survey of lens
fields, and here we present results for the first eight systems
that we have targeted for multi-object spectroscopy.  We characterize
the environment within a $\sim\!6\arcmin$ diameter field around each
of the eight lenses, and quantify how those environments affect the
lens potentials.

Going beyond the lenses' immediate environments, we also consider
the degree to which massive structures along the line of sight to a
lens affect the lens potential.  The prevalence and importance
of interloping structures in lens fields is poorly understood.
Observationally, there appear to be bound groups along the lines
of sight to B0712+472 \citep[10 members;][]{fl} and MG 1131+0456
\citep[3 members;][]{T&Kmg1131-00}.  Overdensities of galaxies are
seen in the fields of several other strong lenses \citep{faure,
morgan}, but it is not yet known whether they indicate massive
bound structures, and whether any such structures are associated
with the lens galaxies or lie elsewhere along the lines of sight.
On the theoretical side, studies have yielded conflicting results
as to whether line-of-sight structures are very important or
negligible for lensing
\citep[e.g.,][]{seljak-94,barkana-96,keeton-97,premadi}.  We show
here that lenses are, in principle, sensitive to structures over
a wide range of redshifts and projected spatial offsets, so the
practical importance of interloping structures depends only on
how common they are.  Our photometric/spectroscopic pencil-beam
survey of lens fields enables us to self-consistently identify
any prominent structures at all relevant redshifts, and to assess
their actual contributions to observed lenses.

Separate from lensing, an important by-product of our survey is a
sizable sample of poor groups at intermediate redshifts.  Only a
few such samples are presently known \citep{carlberg,wilman,gerke}.
Groups are important laboratories for studies of galaxy evolution
because they are the most common environments for galaxies, and
are also relatively simple systems in which the range of mechanisms
thought to drive galaxy evolution (primarily galaxy--galaxy
interactions) is much narrower than in hotter, denser clusters
\citep[][hereafter ZM98]{zabmulch1}.  Unfortunately, poor groups
are notoriously difficult to identify using conventional methods
for finding clusters, due to their low projected surface densities,
faint X-ray luminosities, and inefficiency for weak lensing.
The velocity dispersions of nearby groups range from
$\sigma_r \sim 200$ \kms\ for systems that are X-ray faint,
late-type dominated, dynamically young, and generally similar to
the Local Group; to $\sigma_r \sim 300$--500 \kms\ for systems that
are richer, X-ray luminous, early-type dominated, and dynamically
more evolved; up to $\sigma_r \sim 1000$ \kms\ for rich clusters.
In nearby X-ray luminous groups with $\sigma_r \sim 300$--500 \kms,
there is always a giant elliptical that lies at the center of the
group potential, which suggests that such galaxies form in groups
via interactions prior to being accreted by rich clusters (ZM98).
Groups at intermediate redshifts like those we describe here will,
in conjunction with nearby group samples, permit us to observe the
evolution of groups directly.

The organization of this paper is as follows.  In \S 2 we describe
our sample of eight lens systems and summarize previous work on them.
In \S 3 we present our spectroscopic data in the eight lens fields.
In \S 4 we determine the membership, kinematics, and centroids of
the groups, and use those properties to quantify how the
environments affect the lens models.  We also explore the effects
of line-of-sight structures on the lens models.  We summarize our
results and conclusions in \S 5.  We present the formalism for
computing the convergence and shear arising from perturbing
structures anywhere along the line of sight in an Appendix.
Where necessary, we assume a cosmology with $\Omega_M = 0.3$,
$\Omega_\Lambda = 0.7$, and $H_0 = 70$ km s$^{-1}$ Mpc$^{-1}$.

\section{The Sample}

Our sample consists of eight known gravitational lens systems with
lens galaxies at intermediate redshifts $0.25<z_l<0.5$.  Four of
the lenses (MG 0751+2716, PG 1115+080, B1422+231, and MG 1654+1346)
were suspected from previous studies to have complex environments.
We chose the other four lenses (BRI 0952$-$0115, PMN J2004$-$1349,
B2114+022, and HE 2149$-$2745) because of the availability of prior
imaging and photometry as well as accessibility from Las Campanas
Observatory.  In the remainder of this section we briefly review
prior studies of these eight lenses.  The data from this section are
summarized in Table \ref{lenssample}.

\begin{deluxetable*}{lllllrccccc}
\tablecolumns{11}
\tabletypesize{\footnotesize}
\tablecaption{Gravitational Lens Galaxies\label{lenssample}}
\tablehead{
\colhead{Lens} & \colhead{RA\tablenotemark{b}} & \colhead{Dec\tablenotemark{b}} & \colhead{$z_l$} & \colhead{I\tablenotemark{b}} & \colhead{$z_s $\tablenotemark{b}} & \colhead{$\Delta t$\tablenotemark{e}} & \colhead{Images\tablenotemark{b, f}} & \colhead{$kT_{X}$\tablenotemark{e}} & \colhead{$N_{\rm grp}$\tablenotemark{d, e}} & \colhead{$\sigma_r$\tablenotemark{e}}\\
& \multicolumn{2}{c}{(J2000)} & & \colhead{[mag]} & & \colhead{[days]} & & \colhead{[keV]} & & \colhead{[\kms]} }
\startdata
MG 0751+2716    & 07:51:41.46 &   +27:16:31.4 & 0.349\tablenotemark{a} &  21.26 & 3.20 &  -          & R   & -                  & 2  & -      \\
BRI 0952$-$0115 & 09:55:00.01 & $-$01:30:05.0  &(0.41)\tablenotemark{c} &  21.21 & 4.50 &  -          & 2   & -                    & -  & -    \\
PG 1115+080     & 11:18:17.00 &   +07:45:57.7  & 0.31\tablenotemark{b}  &  18.92 & 1.72 & 25.0$\pm$2.0& 4   & 0.8$\pm$0.2          & 4  & 270$\pm$70 \\
B1422+231       & 14:24:38.09 &   +22:56:00.6  & 0.34\tablenotemark{b}  &  19.66 & 3.62 &  -          & 4   & 1.0$^{+\inf}_{-0.3}$ & 5  & 550$\pm$50 \\
MG 1654+1346    & 16:54:41.83 &   +13:46:22.0  & 0.254\tablenotemark{a} &  17.9  & 1.74 &  -          & R   & -                    & -  & -          \\
PMN J2004$-$1349& 20:04:07.07 & $-$13:49:30.7  & -                      & -      & -    &  -          & 2   & -                    & -  & -          \\
B2114+022       & 21:16:50.75 &   +02:25:46.9  & 0.316\tablenotemark{a} &  18.63 & -    &  -          & 2+2 & -                    & -  & -          \\
                &             &                & 0.59\tablenotemark{b}  &        &      &             &     &                      &    &            \\
HE 2149$-$2745  & 21:52:07.44 & $-$27:31:50.2  & 0.50\tablenotemark{b}  &  19.56 & 2.03 & 103.0$\pm$12.0& 2 & -                    & -  & -          \\
\enddata
\tablenotetext{a}{Data from this work}
\tablenotetext{b}{Data from references in the text and from the CASTLES website (\url{http://cfa-www.harvard.edu/castles/}).}
\tablenotetext{c}{Photometric redshift \citep{kochanekfundplane-00}}
\tablenotetext{d}{Number of previously known group members in addition to the lens galaxy.}
\tablenotetext{e}{References in text.}
\tablenotetext{f}{R means an Einstein ring.}

\end{deluxetable*}

MG 0751+2716 (hereafter MG0751), discovered as a part of the
MIT--Greenbank--VLA search for gravitational lenses, is a radio lens
with four images and a partial ring \citep{Leharmg0751-93}. Optical
imaging of the system by \citet{Leharmg0751-97} identified an
$R = 21.3$ galaxy (G3) located $0.2\arcsec$ northeast of the
brightest radio spot as the likely lens galaxy.  G3 is a satellite
of a much brighter $R=19.1$ galaxy (G1) located $6\arcsec$ away.
\citet{T&Kmg0751-99} determined the redshifts of the galaxies to
be $z_{G1}=0.3501\pm0.0003$ and $z_{G3}=0.3502\pm0.0003$.  They
also found a nearby emission line galaxy to have redshift
$0.3505\pm0.0003$, indicating that the lens galaxy lies in a small
group with at least three members.  Lens models by
\cite{Leharmg0751-97} suggest that MG0751 requires more external
shear that can be accounted for by the observed galaxies, which is
consistent with the hypothesis that the lens environment is complex.
The redshift of the source quasar is $z_s = 3.200\pm0.001$
\citep{T&Kmg0751-99}.

BRI 0952$-$0115 (hereafter BRI0952) was discovered by
\citet{mcmahonbri0952-92} as a doubly imaged $z_s=4.5$ optical
quasar.  The quasar is also detected at millimeter wavelengths
\citep{omontbri0952-96}.  \citet{keetonoptical-98} found that
the lens is a flattened early type galaxy, and
\citet{kochanekfundplane-00} estimated a lens redshift of
$z_l=0.41\pm0.05$ based on fundamental plane fitting.  Because the
separation between the images is small ($0.9\arcsec$) and the lens
galaxy is faint \citep[21.9 in F675W;][]{keetonoptical-98}, the lens
redshift has not been determined spectroscopically.

PG 1115+080 (hereafter PG1115) is a lens system discovered by
\citet{weymannpg1115}, in which a radio-quiet quasar at redshift
$z_s=1.722$ is lensed into four images \citep{hegepg1115}.  The
lens galaxy was first detected by \citet{hhpg1115}; its redshift
was estimated by \citet{angoninpg1115}, and later improved by
\citet{kundicpg1115} and \citet{tonrypg1115} to
$z_l=0.3098\pm0.0002$.  \citet{youngpg1115} suggested the presence
of a small group of galaxies near the lens.  This was confirmed by
\citet{kundicpg1115} and \citet{tonrypg1115}, who measured the
redshifts of a total of four galaxies within $20\arcsec$ of the
lens galaxy.  \citet{kundicpg1115} estimated a group velocity
dispersion of $\sigma_r = 270 \pm 70$ \kms\ from four galaxies,
while \citet{tonrypg1115} estimated $\sigma_r = 326$ \kms\ from
a slightly different set of four galaxies. \citet{grantb1422-04}
detected diffuse X-ray emission that is associated with the group
and that has a temperature $kT \sim 0.8\pm0.2$ keV; this value is
consistent with typical values for low-redshift poor groups, but
somewhat high given the measured group velocity dispersion and
the local $\sigma_r$-$T_X$ relation
\citep[][hereafter MZ98]{zabmulch2}.  PG1115 is one of nine known
strong lens systems for which the time delay between different
images has been measured, so it can be used to determine $H_0$.
\citet{schechterpg1115} measured the light curves of the different
images and estimated the time delays, and \citet{barkanapg1115}
then reanalyzed the data to give more precise results: the delay
between images B and C is $t_{BC}=25.0_{-3.8}^{+3.3}$ days, and
the ratio of the delays between A (actually a combination of the
close images A$_1$ and A$_2$), B, and C is
$t_{AC}/t_{BA} = 1.13_{-0.17}^{+0.18}$.  PG1115 is one of the
lenses with ``anomalous'' flux ratios thought to indicate some
sort of small-scale structure in the lens galaxy
\citep[e.g.,][]{MM,chiba,DK,foldreln}.

B1422+231 (hereafter B1422) is a four-image lens discovered by  
\citet{patnaikb1422-92} while searching for small-separation
lenses among flat spectrum radio sources in the Jodrell
Bank--VLA Astrometric Survey \citep[JVAS;][]{patnaik-92,
browne-98,wilkinson-98,kingb2114-99}.  The source is a radio
loud quasar at $z_s=3.62$ \citep{patnaikb1422-92}, and the lens
is a luminous elliptical at $z_l=0.3374$ \citep{impeyb1422-96,
kundicb1422-97}.  The lens galaxy and five nearby galaxies
form a group at $z_g = 0.338$ with a rest-frame line-of-sight
velocity dispersion of $\sigma_r = 550 \pm 50$ \kms\
\citep{kundicb1422-97}.  Lens models for B1422 require a
significant shear $\gamma\sim 0.20$--0.26, which may be
attributable to the group environment \citep{hoggb1422,
keeton-97,dobler}.  Indeed, from their estimate of the group's
velocity dispersion and centroid, \citet{kundicb1422-97}
estimated $\gamma=0.23$ and pointed out that the group will
also create some convergence $\kappa$ that may affect the lens
potential.  \citet{grantb1422-04} detected B1422 in X-rays
(0.5--2 keV) and determined a temperature of
$kT=1.0^{+\infty}_{-0.3}$ keV, which is consistent with the
value expected for a poor group (MZ98).  B1422 is another lens
with ``anomalous'' flux ratios \citep{mao,chiba}. 

MG 1654+1346 (hereafter MG1654) was originally detected in the
MIT--Greenbank--VLA survey.  \citet{langstonmg1654-88,langstonmg1654-89}
recognized its unusual structure in a VLA snapshot and obtained
radio and optical mapping.  The source is a $z_s=1.74$ radio quasar
with a compact core and two extended radio lobes.  The southwest
lobe is lensed into a ring by a $z_l=0.254$ giant elliptical galaxy
\citep{langstonmg1654-88,kochanekfundplane-00}.
\citet{langstonmg1654-89} noted an enhancement of the number density
of galaxies near the lens; some of the nearby galaxies are comparable
in brightness to the lens galaxy, suggesting a complex environment.

PMN J2004$-$1349 (hereafter PMN2004) is a two-image lens discovered
in a search for radio lenses in the southern sky \citep{winnpmn2004-01}.
The radio spectral index of the images is typical for radio-loud
quasars, so the source is considered to be a quasar despite the lack
of an optical spectrum and a measured redshift \citep{winnpmn2004-01}.
Based on photometry, \citet{winnpmn2004-01} suggested a lens redshift
in the range $0.5<z_l<1.0$.  Higher-resolution imaging by
\citet{winnpmn2004-03} revealed a spiral lens galaxy (only the fifth
one known) and showed that the color differences between the two
images at optical and near-infrared wavelengths can be explained by
differential extinction.  The extinction analysis can be used to
infer the lens redshift; it seems to imply somewhat low values
($0.03 \lesssim z_l \lesssim 0.36$), but that result depends on
assumptions about the extinction curve.
 
B2114+022 (hereafter B2114) was discovered as part of the search
for lenses in JVAS \citep{kingb2114-99}.  Radio maps show four
distinct components within $2.4\arcsec$ in a configuration that
is atypical for lenses.  Furthermore, the sources can be divided
into two pairs with distinct radio surface brightnesses and radio
spectra: sources A and D are similar to each other, and sources
B and C are similar to one another, but the two pairs are clearly
different.  Ground-based and HST optical imaging and spectroscopy
do not detect the lensed images but reveal two lens galaxies at
$z_{l1}=0.3157$ and $z_{l2}=0.5883$ \citep{augustob2114-01},
suggesting a complex lensing geometry.  \citet{chaeb2114-01} could
explain two of the radio components (A and D) as lensed images
using a two-plane lens model.  It is not known whether the other
components (B and C) are images of the same source (unlikely),
images of a different source, or structure related to the G1 lens
galaxy.  No lens models that explain these components have been
published.

HE 2149$-$2745 (hereafter HE2149) is a doubly imaged broad absorption
line quasar at redshift $z_s=2.033$, which was discovered by
\citet{wisotskihe2149-96} in the Hamburg/ESO wide-angle survey for
bright quasars.  \citet{burudhe2149-02} reported a redshift of
$z_l=0.495\pm0.01$ for the elliptical lens galaxy, consistent with
the photometric redshift estimate of $z_l=0.43_{-0.06}^{+0.07}$
\citep{kochanekfundplane-00}.  \citeauthor{burudhe2149-02} also
measured the time delay between the two images to be
$\Delta t = 103\pm12$ days.  Based on the large number of red
non-stellar objects in R-band images of the field around the lens,
\citet{lopezhe2149-98} suggested that the lens galaxy might be
a member of a cluster.

\section{Spectroscopic Data}

We first identified galaxies for follow-up spectroscopy from two-color,
wide-field imaging of each lens field.  We obtained deep images in
I and either V or R during the period from May 2002 to June 2004
using the $36\arcmin\times36\arcmin$ Mosaic Imager on the 4-m
telescopes at Kitt Peak National Observatory and Cerro Tololo
Inter-American Observatory.  We reduced these images and extracted
photometric catalogs following standard methods using
IRAF\footnote{IRAF is distributed by the National Optical Astronomy
Observatories, which are operated by the Association of Universities
for Research in Astronomy, Inc., under cooperative agreement with
the National Science Foundation.} and SExtractor \citep{bertin}.
A more detailed description of the photometric analysis is presented
in a separate paper \citep{kurtis}.

We selected spectroscopy targets using a prioritization scheme
based on objects' colors and projected distances from the lens.
Highest priority was given to objects populating a red sequence
in the color--magnitude diagram that is consistent with the lens
redshift, and to targets that lie within a group-like virial radius
of   0.7 Mpc (ZM98), which corresponds to $\sim\!3\arcmin$ over
the redshift range of our lenses.  We applied a magnitude cut at
$I = 21.5$ to assure reasonable exposure times.  This limiting
magnitude corresponds to $I^{*}+4.2$ at the low redshift limit of
our sample ($z = 0.25$) and to $I^{*}+2.8$ at the high redshift
limit ($z=0.5$), where $I^{*}$ is the observed magnitude of an
$L_{*}$ galaxy, adopted from \citet{kurtis}.  We obtained
multislit spectroscopy during two observing runs, March 1--4 and
August 30--September 2, 2003, with the Low Dispersion Survey
Spectrograph \citep[LDSS-2;][]{ldss} at the 6.5-m Magellan 2
(Clay) telescope at Las Campanas Observatory.  All spectra were
taken with the medium blue grism (300 l/mm; 5000\AA\ blaze) over
a wavelength range of 3900--8000\AA.  We used $1.03\arcsec$
slitlets, resulting in a spectral resolution of $\approx 15$\AA\
FWHM.  Each slitmask had dimensions $\sim5\arcmin\times7\arcmin$,
typically contained 20--30 targets, and was observed for
4$\times$900 s.

\begin{figure*}
\plotone{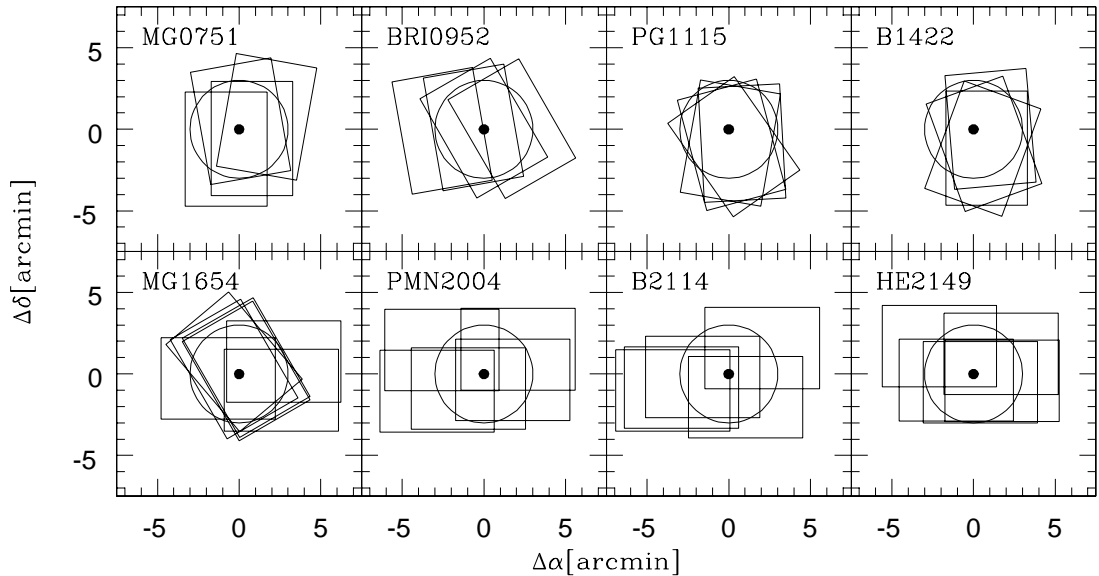}
\caption{
Sky coverage for LDSS-2 multislit spectroscopy in the fields of the
eight lens systems in our sample.  Each panel is
$15\arcmin\times15\arcmin$, centered on the lens galaxy.  Rectangles
show the positions of our slit masks.  Each slitmask covers
$\sim5\arcmin\times7\arcmin$ and includes 20--30 slitlets parallel
to the long side of the mask.  The $3\arcmin$ circle around the
lens galaxy corresponds to $\sim 0.7$ Mpc, a group-like virial at
the redshifs of the lenses in our sample.  We observed four masks for
each of MG0751, BRI0952, PG1115, and B1422, five masks for each of
PMN2004, B2114, and HE2149, and seven masks for MG1654.  The masks
from the first observing run lack a particular sampling pattern but
provide excellent coverage of the expected group virial radius in all
cases except B1422, where a small portion remains unsampled.  The
masks from the second observing run were tiled in an attempt to
maximize the sky coverage while still providing good sampling of the
immediate surroundings of the lens.
\label{coverage}}
\end{figure*}

Our sky coverage is shown in  Figure \ref{coverage}.  Each panel
spans $15\arcmin\times15\arcmin$ and is centered on the lens galaxy.
Rectangles show the outlines of our slitmasks.  There is a noticeable
difference in the sampling between the two observing runs.  In March
2003, when we observed MG0751, BRI0952, PG1115, B1422, and MG1654,
the masks lay mostly on top of one another, providing exhaustive
coverage of the immediate surroundings of each lens galaxy.  Because
most of our fields were far north for Magellan, we aligned the masks
with the paralactic angle to minimize the effects of atmospheric
dispersion.  In contrast, the masks for MG1654, PMN2004, B2114, and
HE2149 in the August 2003 observing run were tiled to cover a larger
area of the sky but still overlap significantly in the $3\arcmin$
projected radius around the lens.  No attempt was made to align the
masks with the paralactic angle because the targets were close to
or south of the celestial equator and were therefore observed at
relatively low airmass.

The figure shows that our sampling is very good within $3\arcmin$
around the lens for all fields, with a minor exception in B1422.  We
observed four masks for each of MG0751, BRI0952, PG1115 and B1422,
five masks for each of PMN2004, B2114 and HE2149, and seven masks
for MG1654 (which was observed during both runs).  We discuss our
spectroscopic completeness below.

We reduced all spectra using standard IRAF procedures and corrected
them to the local standard of rest using the IRAF routines RVCORRECT
and DOPCOR.  We determined the radial velocities using the
cross-correlation of absorption lines (XCSAO) and/or using emission
line identifications (EMSAO)\footnote{XCSAO and EMSAO  are routines
in the RVSAO IRAF package \citep{rvsao}}.  If both emission and
absorption line velocities were found, the quoted value is a weighted
average of the two. Marc Postman kindly provided template galaxy
spectra \citep[][and private communications]{templates}.  We visually
inspected every fit to ensure accuracy.

The number of objects observed in each field is listed in
Table \ref{summary}.  Galaxies represent $\sim$42\% of the objects
targeted spectroscopically, while stars originally misclassified
as galaxies are $\sim$22\%.  ``Failed'' targets ($\sim\!32$\%) are
those for which were unable to obtain velocities.  The majority of
failed targets, especially at fainter magnitudes, were absorption-line
systems for which the signal-to-noise was too low to allow successful
cross-correlation.  Other causes of failures are low surface brightness
or poor astrometry (more problematic in March than August).  The
fraction of stars was significantly lower in August (15\%) than in
March (26\%), thanks to improvements in star-galaxy separation made
between the two runs \citep[see][]{kurtis}.  The large number of
stars in PMN2004 is due to the high stellar density at this relatively
low galactic latitude and longitude ($l = 28\degr$, $b=-22\degr$).

\begin{deluxetable*}{lrrrrr} 
\tablecolumns{3} 
\tablecaption{LDSS-2 Observations \label{summary}} 
\tablehead{ \colhead{Lens}    & \colhead{Date}    & \colhead{\# Galaxies\tablenotemark{a}} & \colhead{\# Stars} & \colhead{\# Failed} & \colhead{Total} }
\startdata 

MG0751   &  March 2003 & 38 & 24 & 27 &  89 \\ 
BRI0952  &  March 2003 & 47 & 24 & 19 &  90 \\ 
PG1115   &  March 2003 & 47 & 28 & 11 &  86 \\ 
B1422    &  March 2003 & 53 & 14 & 26 &  93 \\ 
MG1654   &  March 2003 & 39 & 30 & 32 & 101 \\ 
         & August 2003 & 20 & 6  & 35 &  61 \\ 
PMN2004  & August 2003 & 41 & 43 & 35 & 119 \\
B2114    & August 2003 & 38 & 1  & 46 &  85 \\ 
HE2149   & August 2003 & 41 & 8  & 66 & 115 \\ 

\enddata 
\tablenotetext{a}{Including QSOs.}
\end{deluxetable*}

\begin{figure*}
\plotone{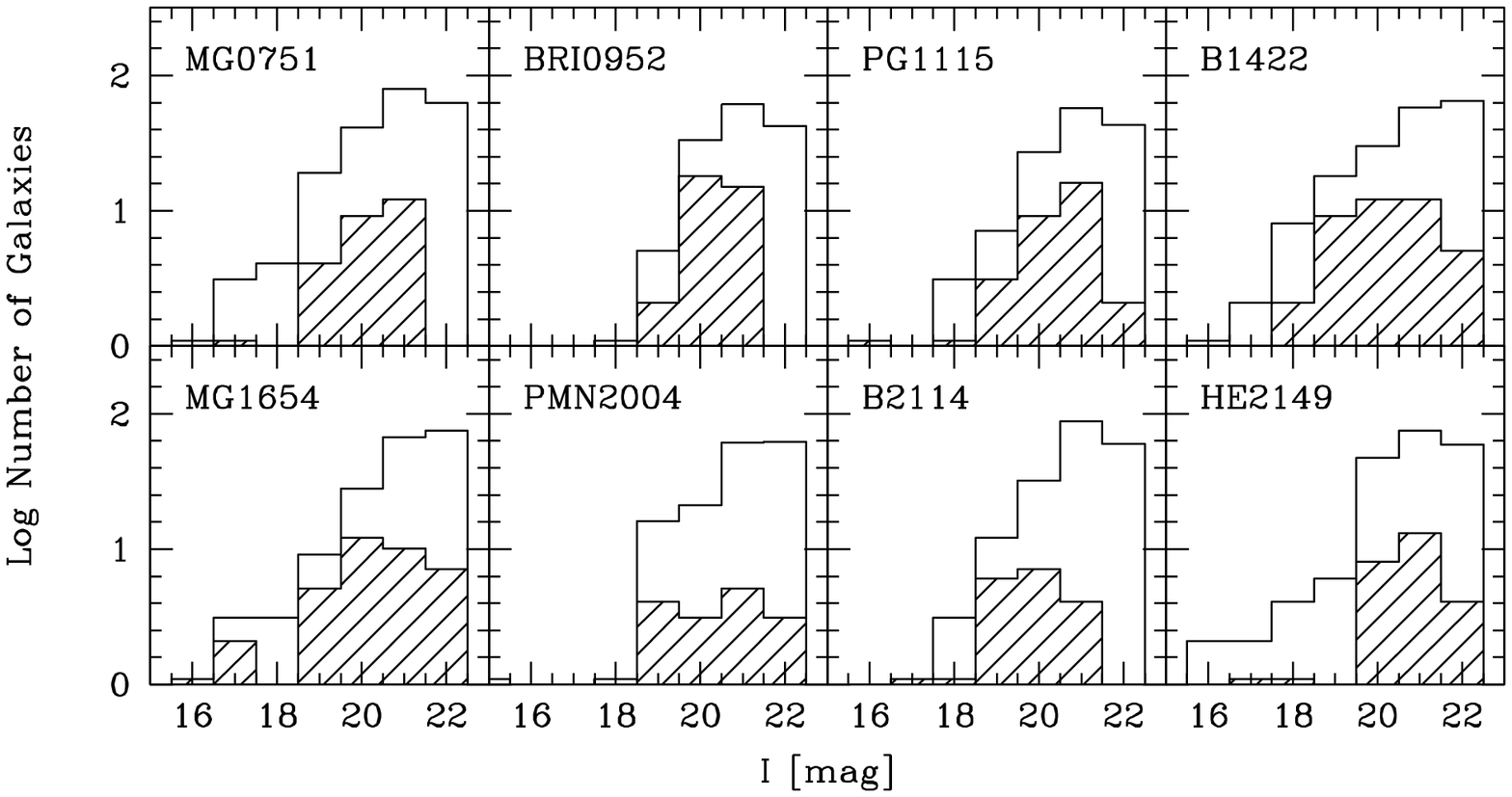}
\caption{
Apparent magnitude histograms of galaxies in the photometric catalog
({\em solid line}), and of those galaxies within $3\arcmin$ of
the lens for which redshifts were obtained ({\it shaded}).
Our target selection scheme, based on colors and projected offsets
from the lens, misses a few of the brightest galaxies.  We miss a
larger fraction of objects at the faint end, but these are presumably
less massive galaxies that do not contribute significantly to the
lens potential.
\label{completeness1}}
\end{figure*}

Figure \ref{completeness1} shows the completeness of our spectroscopy
with respect to the photometry.  The solid-line histogram shows the
magnitude distribution for all galaxies in the photometric catalog
projected within $3\arcmin$ of the lens, while the shaded histogram
shows the subset for which we determined velocities.  Our target
selection scheme, based on colors and projected offsets from the
lens, misses a few of the brightest galaxies (many of which are
likely to be foreground objects).  We miss a larger fraction of
objects at the faint end, but these are very few and/or less massive
galaxies, which would significantly affect the lens potential only
if very close to the lens galaxy.  In total, there are only four
galaxies within $10\arcsec$ of the lens galaxies (the zone in which
a small perturber can have even a moderate effect; see Fig.~\ref{sigma}
below) that are present in our photometric catalog but have no
determined velocities.  Two of those (in B2114 and PMN2004) are
below our spectroscopic magnitude limit of $I=21.5$.  The other
two galaxies are in the field of HE2149, have I magnitudes of
19.97 and 21.08, and lie 7$\arcsec$ and 10$\arcsec$ respectively
from the lens.  In color--magnitude space \citep{kurtis} they lie
on red sequences identified as line-of-sight structures at $z=0.45$
and $z=0.60$ (see \S 4.5).

We were never able to put slits on all of our highest-priority
objects; this limitation is inherent to multislit spectroscopy.
For lensing purposes, the main effect of spectroscopic incompleteness
is to cause us to {\em underestimate} environment-related lensing
biases (see \S\S 4.4--4.5).  In particular, the fact that we
prioritize galaxies thought to lie at the lens redshift (color
selection) means that we undersample line-of-sight structures and
hence underestimate their contributions to the lens potential.

As already mentioned, we improved our photometric catalogs after some
of the spectroscopic targets were selected.  As a result, there are
15 galaxies whose velocities we measured that do not actually appear
in the final photometric catalog.  (These galaxies lie under bleed
trails in the imaging, so accurate photometry is not possible.)  In
addition, there are another 25 galaxies whose velocities we measured
that are (mis-)classified in the final photometric catalog as stars.
We omit all of these galaxies when comparing the spectroscopic and
photometric catalogs for the purpose of understanding our spectroscopic
completeness (i.e., they are excluded from Fig.\ \ref{completeness1}).
However, these remain a part of our spectroscopic catalog and all
analyses based on that catalog.

To estimate the zero-point velocity correction and external velocity
errors, we cross-correlate 403 sky spectra extracted from our data
with the same templates used for the galaxy spectra and find a mean
velocity of $\bar{\upsilon} = 40 \pm 50$ \kms.  We also
determine the velocities of 153 of the serendipitously observed stars
and find $\bar{\upsilon} = 30 \pm 180$ \kms.  Both methods give
mean velocities comparable to or smaller than the dispersion, and
much smaller than the velocities of the objects in the sample.  We
therefore conclude that no zero point correction is needed.

Table \ref{catalog} lists our spectroscopic catalog.  For each entry
we give the catalog name, J2000 coordinates calibrated to USNO-B2.0,
projected distance from the lens in arcmin, aperture magnitude within
a fixed physical size of $\sim$6.5 kpc \citep[see][]{kurtis}, the
heliocentric radial velocity,  velocity error, redshift and redshift
error.  The last column describes the method from which the velocity
was obtained: 1 for absorption lines, 2 for emission lines, or 3 for
a combination of both.  Missing data means that the object was not
in the final photometric catalog as described above.  Such objects
have no identification names and are numbered successively starting
with 90001. 

\begin{deluxetable*}{lrrrrrrrrc}
\tabletypesize{\footnotesize}
\tablecolumns{10}
\tablecaption{Lens Field Galaxy Properties --- EXAMPLE\label{catalog}}
\tablehead{\colhead{ID}
& \colhead{$\alpha$}   & \colhead{$\delta$}    & \colhead{$b$}  & \colhead{I} & \colhead{$cz$}   & \colhead{$\Delta cz$} & \colhead{$z$} &\colhead{$\Delta z$}   & \colhead{Spectral}\\
\colhead{} & \colhead{[hh:mm:ss]}   & \colhead{[dd:mm:ss]}    & \colhead{[']}  & \colhead{[mag]} & \multicolumn{2}{c}{[\kms]}  & \colhead{}  & \colhead{}   & \colhead{Type}}
\startdata
\sidehead{\bf MG0751: MWKZ GAL}
... 9809  &  07:51:30.20  &  27:14:43.1  &  3.10  & 20.9 & 168070 & 30 & 0.56023 & 0.000083 &  2 \\ 
... 9739  &  07:51:30.77  &  27:17:55.7  &  2.76  & 20.6 & 105970 & 70 & 0.35323 & 0.000237 &  2 \\ 
... 9726  &  07:51:31.00  &  27:14:16.7  &  3.24  & 20.5 & 175630 & 50 & 0.58543 & 0.000173 &  3 \\ 
... 9582  &  07:51:32.04  &  27:12:59.2  &  3.30  & 19.0 &  91080 & 30 & 0.30360 & 0.000113 &  3 \\ 
... 9606  &  07:51:32.28  &  27:12:58.9  &  4.10  & 20.8 & 106950 & 100 & 0.35650 & 0.000320 &  1 \\ 
... 9570  &  07:51:32.49  &  27:17:38.6  &  2.29  & 20.2 & 104680 & 30 & 0.34893 & 0.000113 &  3 \\ 
... 9560  &  07:51:32.73  &  27:14:42.1  &  2.67  & 21.0 & 106300 & 30 & 0.35433 & 0.000113 &  2 \\ 
... 9333  &  07:51:34.61  &  27:15:45.5  &  1.71  & 19.5 &  91220 & 50 & 0.30407 & 0.000153 &  1 \\ 
... 9297  &  07:51:34.66  &  27:13:37.7  &  3.28  & 18.4 &  28910 & 40 & 0.09637 & 0.000133 &  1 \\ 
... 9274  &  07:51:35.24  &  27:17:37.1  &  1.76  & 19.6 & 105350 & 110 & 0.35117 & 0.000380 &  1 \\ 
... 9225  &  07:51:35.43  &  27:17:07.9  &  1.48  & 19.0 &  79820 & 60 & 0.26607 & 0.000213 &  3 \\ 
... 9239  &  07:51:35.75  &  27:15:22.8  &  1.72  & 20.9 & 104960 & 80 & 0.34987 & 0.000280 &  1 \\ 
... 9100  &  07:51:36.66  &  27:19:39.8  &  3.31  & 18.4 &  74510 & 50 & 0.24837 & 0.000160 &  1 \\ 
... 9047  &  07:51:36.96  &  27:19:27.9  &  3.10  & 18.3 & 104220 & 40 & 0.34740 & 0.000140 &  2 \\ 
... 9120  &  07:51:36.98  &  27:18:40.6  &  2.37  & 20.9 &  72100 & 60 & 0.24033 & 0.000203 &  2 \\ 
... 9049  &  07:51:37.73  &  27:18:06.5  &  1.78  & 20.4 &  71940 & 50 & 0.23980 & 0.000163 &  2 \\ 
... 8794  &  07:51:37.92  &  27:16:12.9  &  0.85  & 0.0 & 167590 & 100 & 0.55863 & 0.000320 &  1 \\ 
... 9006  &  07:51:38.04  &  27:17:33.7  &  1.28  & 20.1 & 168620 & 90 & 0.56207 & 0.000307 &  1 \\ 
... 8816  &  07:51:40.32  &  27:16:22.1  &  0.31  & 20.9 & 104940 & 100 & 0.34980 & 0.000323 &  1 \\ 
... 8669  &  07:51:41.07  &  27:19:43.7  &  3.20  & 18.9 & 147320 & 120 & 0.49107 & 0.000390 &  1 \\ 
... 8682  &  07:51:41.50  &  27:16:31.9  &  0.00  & 20.1 & 104810 & 120 & 0.34937 & 0.000410 &  1 \\ 
... 8673  &  07:51:41.84  &  27:16:29.2  &  0.09  & 21.5 & 167090 & 10 & 0.55697 & 0.000037 &  2 \\ 
... 8476  &  07:51:43.25  &  27:17:53.3  &  1.41  & 19.8 & 169470 & 210 & 0.56490 & 0.000687 &  1 \\ 
... 7921  &  07:51:43.26  &  27:16:06.3  &  0.58  & 0.0 & 124780 & 30 & 0.41593 & 0.000103 &  3 \\ 
... 8385  &  07:51:44.18  &  27:16:39.7  &  0.61  & 20.7 & 112430 & 200 & 0.37477 & 0.000657 &  1 \\ 
... 8257  &  07:51:45.32  &  27:17:02.4  &  0.99  & 19.4 & 105110 & 60 & 0.35037 & 0.000200 &  1 \\ 
... 8288  &  07:51:45.50  &  27:18:51.4  &  2.49  & 20.4 &  60380 & 100 & 0.20127 & 0.000317 &  1 \\ 
\enddata
\tablecomments{This table is published in its entirety in the
electronic edition.  A portion is shown here for guidance regarding its
form and content.  Objects with missing data were not found in our final
photometric catalog for reasons explained in the text.}
\end{deluxetable*}

\section{Results and Discussion}

\subsection{Group Membership}

The environments of most strong lenses are not well characterized.
Our first goal is to determine whether each lens galaxy lies in
a group or cluster, and if so to identify the other member galaxies.
Even in cases where groups were already identified (MG0751, PG1115,
and B1422), the number of group members known previously ranged from
three to six.  Our deep, wide-field spectroscopic sampling has the
potential not only to find new groups, but also to increase the
membership of known lens groups to the point where robust
determinations of the group velocity dispersions and centroids are
possible.  These are essential for understanding how a group affects
the lens potential, as discussed in \S 4.4. Additionally, a more
complete inventory of the brightest (most massive) group members,
and their contributions to the lens potential, will also greatly
improve lens models.

We present the redshift histograms for all eight of our lens fields
in Figure~\ref{velocity}.  In the left panels, the shaded histograms
include all galaxies that lie within a projected radius of $3\arcmin$
(a group-like virial radius) around the lens, and $N_{\rm tot}$
indicates the number of galaxies in the histogram.  This number
excludes a few high-redshift AGNs that fall outside the range of
the plot.  The histograms include all galaxies with spectroscopic
redshifts both from our sample and from the literature.  In
particular, we have added the following 16 galaxies to our catalog:
G1, G6, and G7 in MG0751 \citep{T&Kmg0751-99}; the lens galaxy GL
as well as Gx in PG1115 \citep{tonrypg1115}; the lens galaxy G1
as well as G2, G3, G4, G6, G8, G9, G10, and Gx in B1422
\citep{kundicb1422-97,tonrypg1115}; the $z_l=0.59$ lens galaxy in
B2114 \citep{augustob2114-01}; and the $z_l=0.495\pm0.01$ lens
galaxy in HE2149 \citep{burudhe2149-02}.  We have not included
 the lens galaxy in BRI0952 (because it only has a photometric
redshift estimate) or in PMN2004 (because no good redshift estimate
exists).  The vertical line shows the position of the lens galaxy
as listed in Table \ref{lenssample}.  In six cases (see
Table \ref{groupproptable}), there is clearly a peak in redshift
space at or near the lens galaxy redshift.
\begin{deluxetable*}{lrrrrrlrrrrlll}
\tablecolumns{14}
\tabletypesize{\footnotesize}
\tablecaption{Group Kinematic Properties\label{groupproptable}}
\tablehead{
\colhead{Lens} & \colhead{$N_{\rm tot}$} & \colhead{$N_{\rm grp}$} & \colhead{$\alpha_{cen}$} & \colhead{$\sigma_{\alpha,cen}$} & \colhead{$\delta_{cen}$} & \colhead{$\sigma_{\delta,cen}$} & \colhead{$\upsilon_{min}$} & \colhead{$\upsilon_{max}$} & \colhead{$\bar{\upsilon}$} & \colhead{$\delta\bar{\upsilon}$} &\colhead{$z$} & \colhead{$\sigma_{r, {\rm grp}}$} & \colhead{$\delta \sigma_{r, {\rm grp}}$} \\
& & & &[\arcsec] & &[\arcsec]& \multicolumn{2}{c}{[$\;\;km\:s^{-1}$]} & \multicolumn{2}{c}{[$\;\;km\:s^{-1}$]} & \colhead{} & \multicolumn{2}{c}{[$\;\;km\:s^{-1}$]}}
\startdata
MG0751  & 39 & 13 & 07:51:40.7 & $\pm 11$ & $+$27:16:53 & $\pm  9$ & 104220 & 105970 & 104980  & $\pm 100$ & 0.3499 & 320                  & $^{+170}_{-110}$ \\
BRI0952 & 44 &  5 & 09:54:56.1 & $\pm 34$ & $-$01:29:58& $\pm 27$ &  125000 & 127000 & 126510  & $\pm  30$ & 0.4217 & 170\tablenotemark{a} & $^{+150}_{-100}$ \\
PG1115  & 48 & 13 & 11:18:16.8 & $\pm 11$ & $+$07:45:36 & $\pm  9$ &  92120 &  93960 &  92970  & $\pm 110$ & 0.3090 & 440                  & $^{+ 90}_{- 80}$ \\
B1422   & 57 & 16 & 14:24:41.0 & $\pm 11$ & $+$22:55:42 & $\pm  9$ & 100640 & 102810 & 101540  & $\pm 130$ & 0.3385 & 470                  & $^{+100}_{- 90}$ \\
MG1654  & 59 &  7 & 16:54:39.3 & $\pm 27$ & $+$13:47:15 & $\pm 20$ &  75390 &  76210 &  75750  & $\pm 100$ & 0.2525 & 200                  & $^{+120}_{- 80}$ \\
B2114   & 38 &  5 & 21:16:51.4 & $\pm 34$ & $+$02:10:59 & $\pm 27$ &  93000 &  95000 &  94240  & $\pm  80$ & 0.3141 & 110                  & $^{+170}_{- 80}$ \\
\enddata 
\tablenotetext{a}{Velocity dispersion calculated in the manner of
\citet{stddiv}, instead of using the statistical bi-weight estimator
of scale \citep{statistical}. See text.}
\end{deluxetable*}

We determine the group membership by applying a pessimistic
3$\sigma$ clipping algorithm \citep[as suggested by][]{3sigmaclip}
to any redshift peak containing a lens galaxy.  This procedure
removes the galaxy most deviant from the mean redshift and
recalculates the mean and velocity dispersion of the distribution.
If the omitted galaxy is more than three new standard deviations
away from the recomputed mean, it is rejected.  This loop is
executed until an omitted galaxy is not rejected.  We use
statistical bi-weight estimators, which are more robust for small
sample sizes than the standard estimators \citep{statistical},
to calculate the location (mean redshift) and scale (velocity
dispersion).  In Figure~\ref{velocity}, $N_{\rm grp}$ is the number
of group members determined by the 3$\sigma$ clipping algorithm.

The right-hand panels in Figure~\ref{velocity} show a cut within
$\pm$5000 \kms\ of the lens galaxy velocity to present a better
view of the lens groups themselves.  The peak height is lower
than in the full histogram because the bin size is smaller, but
groups are still easily recognizable in six cases: MG0751,
BRI0952, PG1115, B1422, MG1654 and B2114.  {\em The groups in
BRI0952, MG1654, and B2114 are new discoveries.  For the
previously known groups, we have increased the number of group
members from three to 13 in MG0751, from five to 13 in PG1115,
and from six to 16 in B1422.}  The projected spatial distributions
of galaxies in the six groups are shown in Figure \ref{skyplots}.

In the case of BRI0952, we suggest that the lens galaxy may belong
to the five-member group at $z_g=0.422$, which is consistent with
its photometric redshift estimate of $z_l=0.41\pm0.05$.  In B1422,
the group seems to consist of two clumps --- one around the lens
galaxy and the other to the northeast of it (see Figure \ref{skyplots}).
\citet{substructure} have found analogous substructure in nearby
groups.  B1422 is our best sampled group (16 members) and we expect
that increasing the membership of the other groups might reveal
similar clumpiness.  In B2114, the group is associated with the
foreground lens galaxy (the first vertical dashed line).  There is
a second peak slightly in front of the group around the lens galaxy,
but the galaxies are significantly offset from the lens on the sky
and thus not important for lensing.  We classify as ``group members''
only the galaxies in the peak at the lens redshift.

Our spectroscopic findings agree well with the expectations set
by our photometry \citep[see][]{kurtis}.  All six lenses where we
find groups show a compelling red sequence at the lens redshift.
In HE2149, the color--magnitude diagram shows a well defined red
sequence corresponding to $z\sim0.28$, where we see a prominent
line-of-sight structure in Figure \ref{velocity}.  In PMN2004,
the lack of compelling structures in either the color--magnitude
diagram or the redshift histogram suggests that there are no
significant structures along the line of sight, although it is
difficult to draw firm conclusions from non-detections (especially
given incomplete spectroscopy).  In any case, the agreement between
our photometric and spectroscopic results reassures us that we
understand the data and their implications.  These results are
discussed further in \citet{kurtis}.

\begin{figure*}
\includegraphics[height=7in,angle=270]{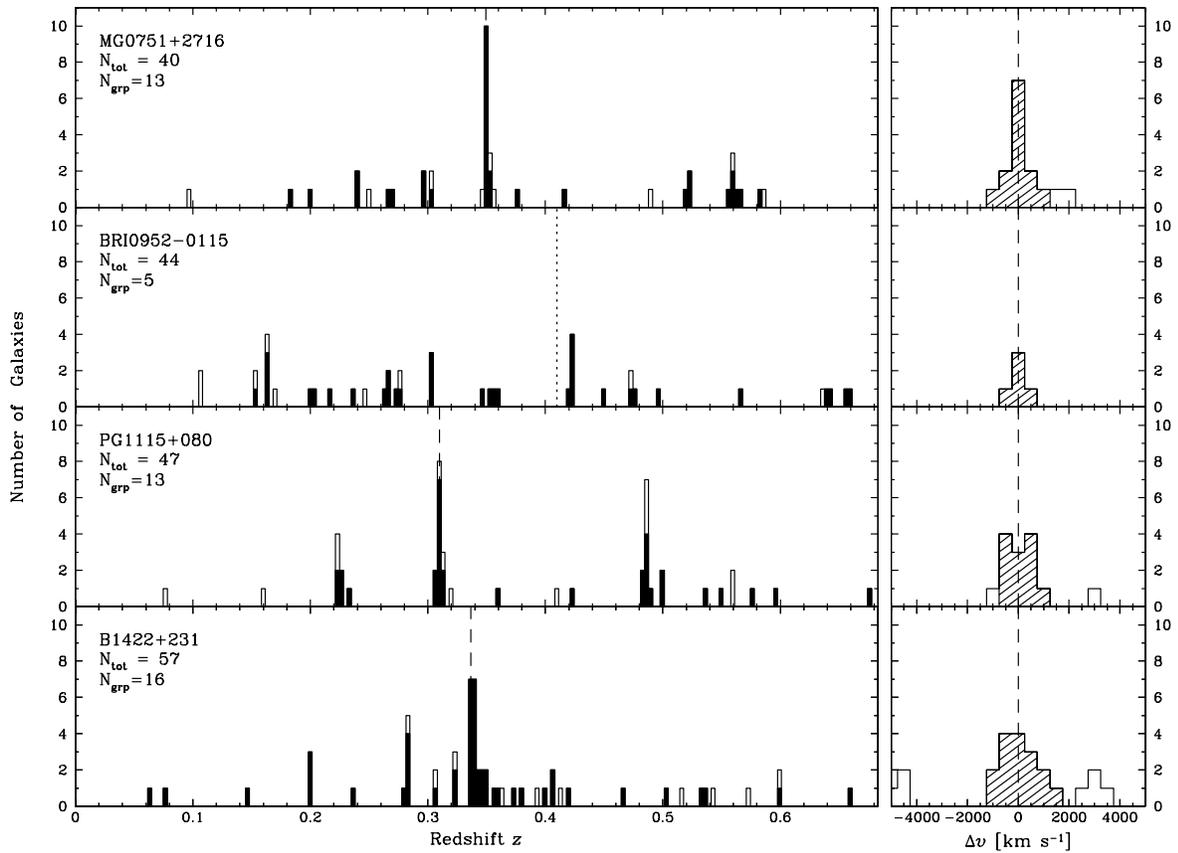}
\caption{\emph{(Left)} Galaxy redshift distributions of
the eight fields in our sample.  The bin size is 1000 \kms.  The
shaded histogram includes all galaxies that lie within a projected
radius of $3\arcmin$ about the lens.  The vertical dashed lines
show spectroscopic lens galaxy redshifts from the literature.
The vertical dotted line for BRI0952 shows a photometric estimate
of the lens galaxy redshift. In PMN2004, the vertical dot-dashed lines
show two different model-implied estimates of the lens galaxy
redshift \citep{winnpmn2004-03}.  $N_{\rm tot}$ is the total number
of galaxies included in the histogram, while $N_{\rm grp}$ is the
total number of group members.  \emph{(Right)} A close-up of the
range $\pm5000$ \kms\ centered on the mean group velocity.  The bin
size is 500 \kms. The shaded histogram shows confirmed group
members.\label{velocity}}
\end{figure*}

\begin{figure*}
\includegraphics[height=7in,angle=270]{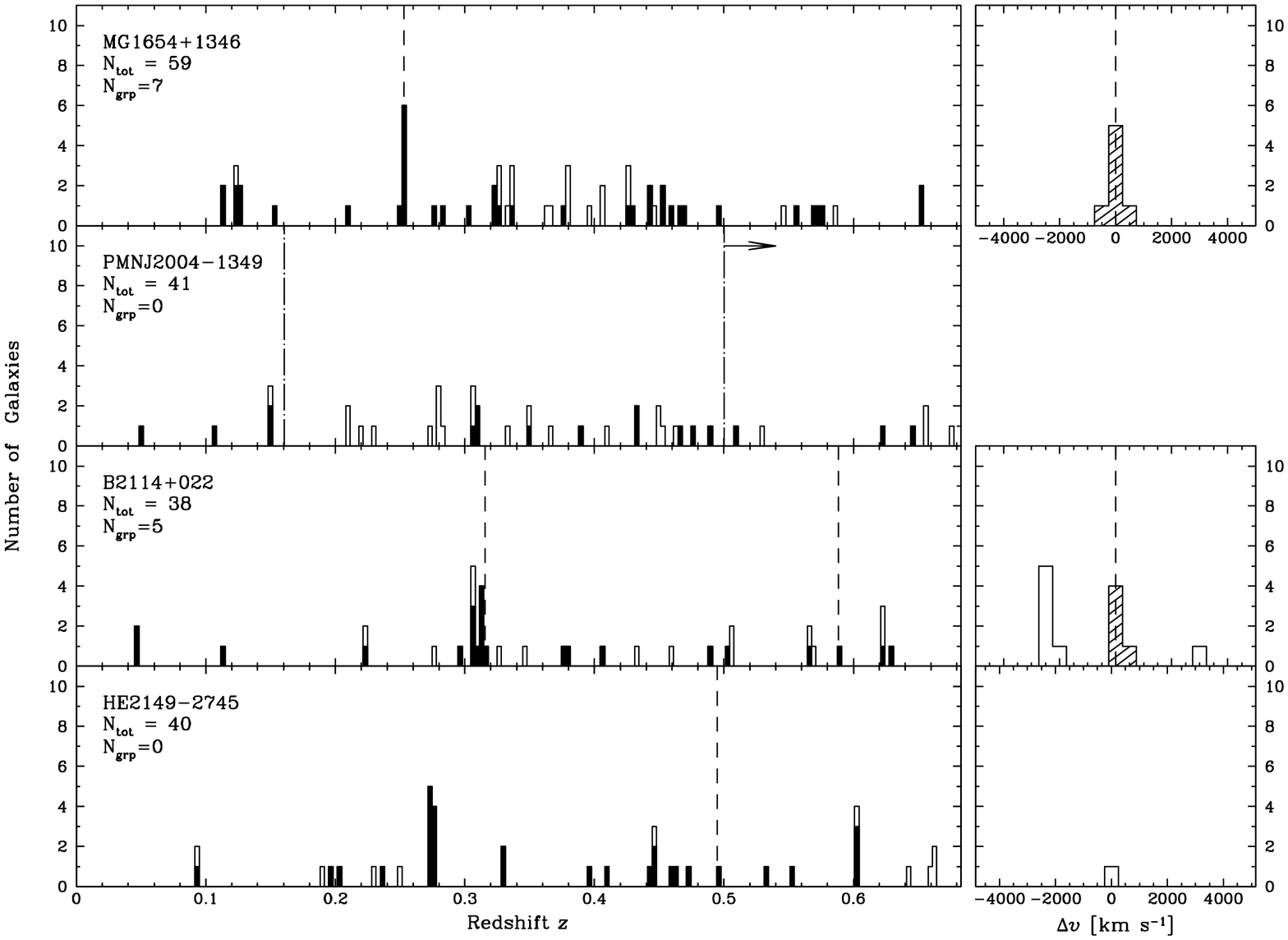}
\figurenum{\ref{velocity}}
\caption{continued.}
\end{figure*}

\subsection{Group Kinematics}

The group velocity dispersion provides a key observable probe of
the group potential and its effect on lens models.  Group velocity
dispersions based on small member catalogs are uncertain and may
be biased, because poor sampling of the underlying velocity distribution
tends to underestimate the true velocity dispersion (ZM98).  Our
more extensive member catalogs now allow us to measure the group
velocity dispersions more accurately than was previously possible.

To determine the mean velocities, $\bar{\upsilon}$, and
line-of-sight velocity dispersions, $\sigma_r$, of the six groups
in our sample, we use bi-weight estimators of location and scale
\citep{statistical} because of their superiority at de-weighting
tails in the velocity distribution.  In BRI0952, which has only
five members, the bi-weight estimator routine fails, so we use the
standard method for calculating the mean velocity and velocity
dispersion from \citet{stddiv}.  In general, standard and bi-weight
methods yield similar means and velocity dispersions.  We use all
known group members, including the 10 found in the literature, to
determine these kinematic properties.  We apply a standard
$1/(1+z_g)$ cosmological correction to the velocity dispersions
\citep{statistical}.  The kinematic properties of the groups are
presented in Table \ref{groupproptable}.

The six groups have velocity dispersions ranging from
$110^{+170}_{-80}$ to $470^{+100}_{-90}$ \kms.  In the nearby
universe, this range of velocity dispersions describes systems
running from dynamically young, unrelaxed systems like the Local
Group, up to more dynamically relaxed, X-ray luminous groups.
We are probably seeing a similar range of groups in our
intermediate-redshift sample.  It is important to note that
the total number of galaxies for which velocities were obtained
is similar in all six fields, and that $N_{\rm grp}$ is roughly
correlated with $\sigma_r$, suggesting that the differences
between the measured $\sigma_r$ values are real and not due
to variable sampling.

As a further check on the accuracy of our group velocity dispersions,
we can determine the X-ray temperature that would be derived by
combining our measured $\sigma_r$ values with the $\sigma_r$-$T_X$
relation for nearby groups and clusters (ZM98), and compare that
with the temperature measured directly from X-ray observations.
Only two of the groups in our sample have been observed in X-rays
with {\em Chandra}: PG1115 and B1422 \citep{grantb1422-04}.  
The X-ray temperatures expected from the $\sigma_r$-$T_X$ relation 
(1.5 keV and 1.7 keV, respectively) are consistent within the 95\%
confidence limit with the observed values (0.8 keV and 1.0 keV,
respectively).  This suggests that our values of $\sigma_r$ for
these groups are reasonable.

\subsection{Group Centroids}

The projected offset of the lens galaxy from the group centroid
is another key ingredient in estimating the contribution of a
group to the lens potential. The position of the brightest group
galaxy relative to the spatial and kinematic centroid of the group
is also an important constraint on models of giant elliptical
formation (ZM98).  (The lens galaxy may or may not be the brightest
group galaxy, as discussed below.)  In this section we calculate
the projected spatial and kinematic centroids of the groups by
averaging the sky positions and velocities, respectively, of all
group members.  For members whose velocities were added from the
literature, we use coordinates from our own photometric catalog
when possible in order to maintain a consistent coordinate system.
Neither the projected spatial centroid nor the mean velocity is
weighted by the luminosity, because we do not want to introduce
an a priori bias toward the brightest group galaxy or assume
implicitly that the mass-to-light ratios for all group members are
the same.  Nevertheless, the luminosity-weighted centroids are
within $2\sigma$ of the unweighted centroids for all groups except
B1422 (where four bright galaxies close to the lens pull the
luminosity-weighted centroid $3.5\sigma$ away from the unweighted
centroid).

\subsubsection{Projected Offset Between the Spatial Centroid and
the Lens Galaxy}

If a group around a lens has a common dark matter halo, the
effects of that halo on the lens potential are sensitive to any
projected spatial offset between the halo centroid and the lens
galaxy.  In particular, the offset determines the relative
importance of convergence (gravitational focusing) and shear
(tidal distortions) from the group halo.  (See \S 4.4 and the
Appendix for details.)  While we obviously want to determine the
offset in each lens/group system, we also seek to understand the
distribution of offsets because that affects the distribution of
convergence and shear, which in turn affects statistical
applications like constraining the dark energy or understanding
the quad/double ratio.  For MG0751, PG1115, and B1422, the new
members we have found allow us to measure the offsets more
precisely.  For the newly discovered groups around BRI0952,
MG1654, and B2114, we are able determine the offsets for the
first time.

\begin{figure*}
\plotone{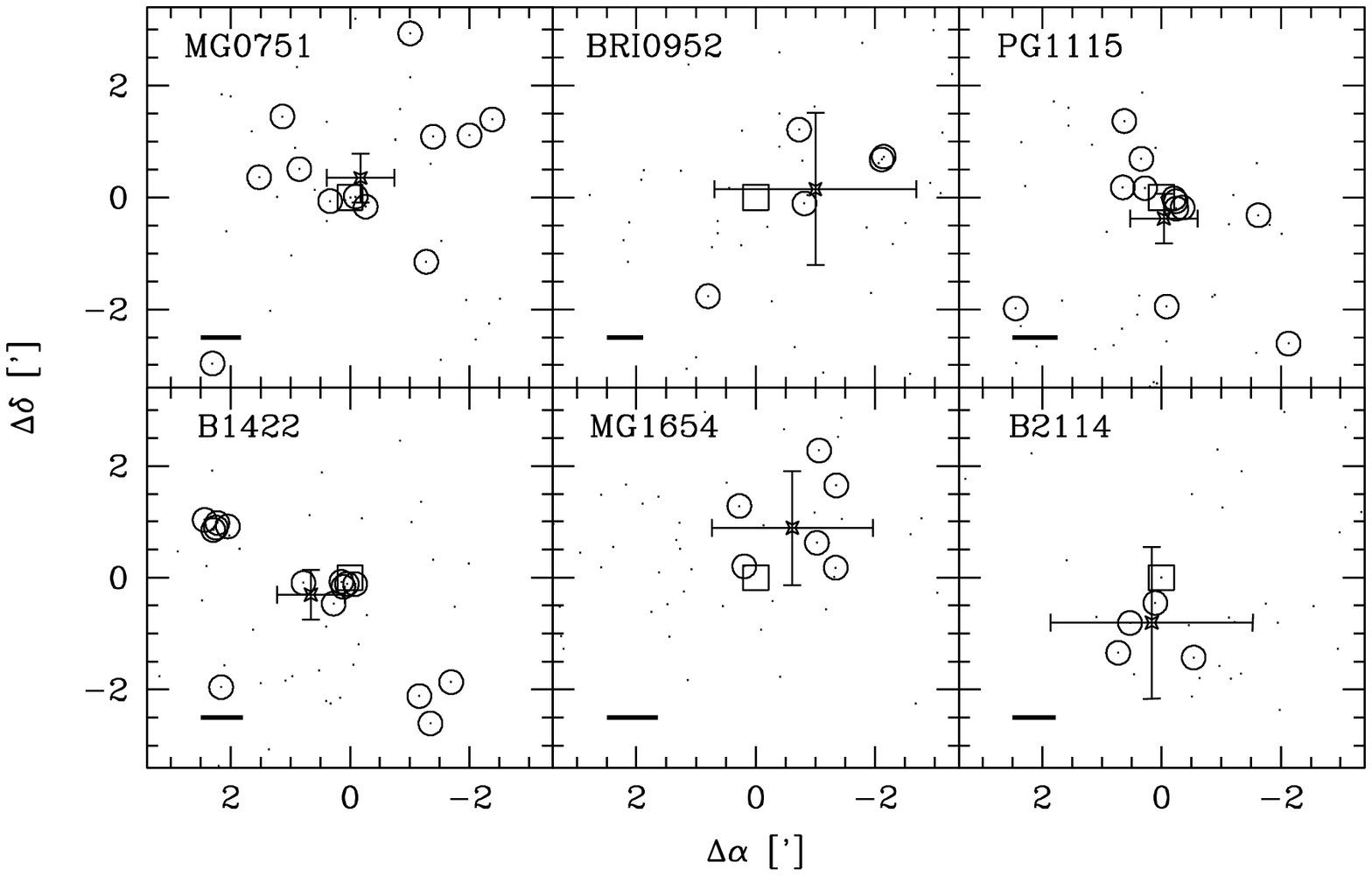}
\caption{
Spatial distribution of the group member galaxies on the sky.  North
is up and east is to the left.  The fields are centered on the lens.
Each panel has an angular size of $6.8\arcmin\times 6.8\arcmin$,
which is roughly equivalent to the typical size of nearby poor groups.
The open circles mark the group galaxies, an open square denotes the
lens galaxy, and a four-pointed star indicates the group centroid
and its $2\sigma$ errorbars (see \S 4.3).  All galaxies with measured
velocities are marked with small solid points.  The scale bar in the
lower left corner of each panel corresponds to 200 kpc at the lens
redshift.\label{skyplots}}
\end{figure*}

Figure \ref{skyplots} shows that in some cases there is a clear
offset between the lens galaxy and the projected group spatial
centroid.  In B1422, the offset is substantial; the lens
galaxy lies outside the $2\sigma$ errors for the group
centroid and is not the galaxy closest to the group centroid.
In MG0751, PG1115 and MG1654 the lens galaxy is only marginally
within the $2\sigma$ centroid errorbars, and is also not the
galaxy closest to the centroid.  We use these spatial offsets in
our calculations of the shear and the convergence due to the lens
environment in \S 4.4.

\subsubsection{Brightest Group Galaxy vs.\ Group Potential}

Key issues that bear on the evolution of groups and their
galaxies are whether the brightest group galaxy (hereafter BGG)
is kinematically and spatially distinct from  (1) the other group
members or (2) from the center of the group potential.  In nearby
X-ray luminous groups ($\sigma_r \gtrsim 300$ \kms), there
is a bright, giant elliptical galaxy that occupies the
center of the potential (as defined by the spatial and kinematic
centroids) and that lies on an orbit distinct from the other group
members (ZM98). Our survey now allows us to ask
the same questions for intermediate-redshift groups, and to
consider what the answers imply about group evolution.

In each of our six groups the BGG appears to have an early-type
morphology. The identifications are as follows:
\begin{itemize}
\item
MG0751: The BGG is the G1 galaxy, which lies $6\arcsec$ from the
lens galaxy and is 2.2 mag brighter in I. \citet{Leharmg0751-97} fit elliptical profiles to G1 and find an acceptable fit. Furthermore, \citet{T&Kmg0751-99} find that G1 is much brighter that the lens galaxy and has a pure absorption line spectrum.
\item
BRI0952: We identify the BGG by visual inspection of our photometry as an elliptical
$2.3\arcmin$ away from the lens galaxy and 1.5 mag brighter in I.
\item
PG1115: The BGG is the giant early-type galaxy labeled G1,
located $12\arcsec$ away from the lens galaxy and 0.7 mag
brighter in I. \citet{impeypg1115} comment on its early type morphology.
\item
B1422: \citet{kundicb1422-97} claim the BGG is the galaxy G3 located $8\arcsec$ from the lens galaxy and 1.5 mag brighter in V. It is the brightest group galaxy in our sample, has a central location, and exhibits an early type spectrum.
\item
MG1654 and B2114: In both cases, the lens galaxies are the brightest galaxies in the groups and are also ellipticals as classified by \citet{csk95} and \citet{augustob2114-01}, respectively.
\end{itemize}

\begin{figure}
\plotone{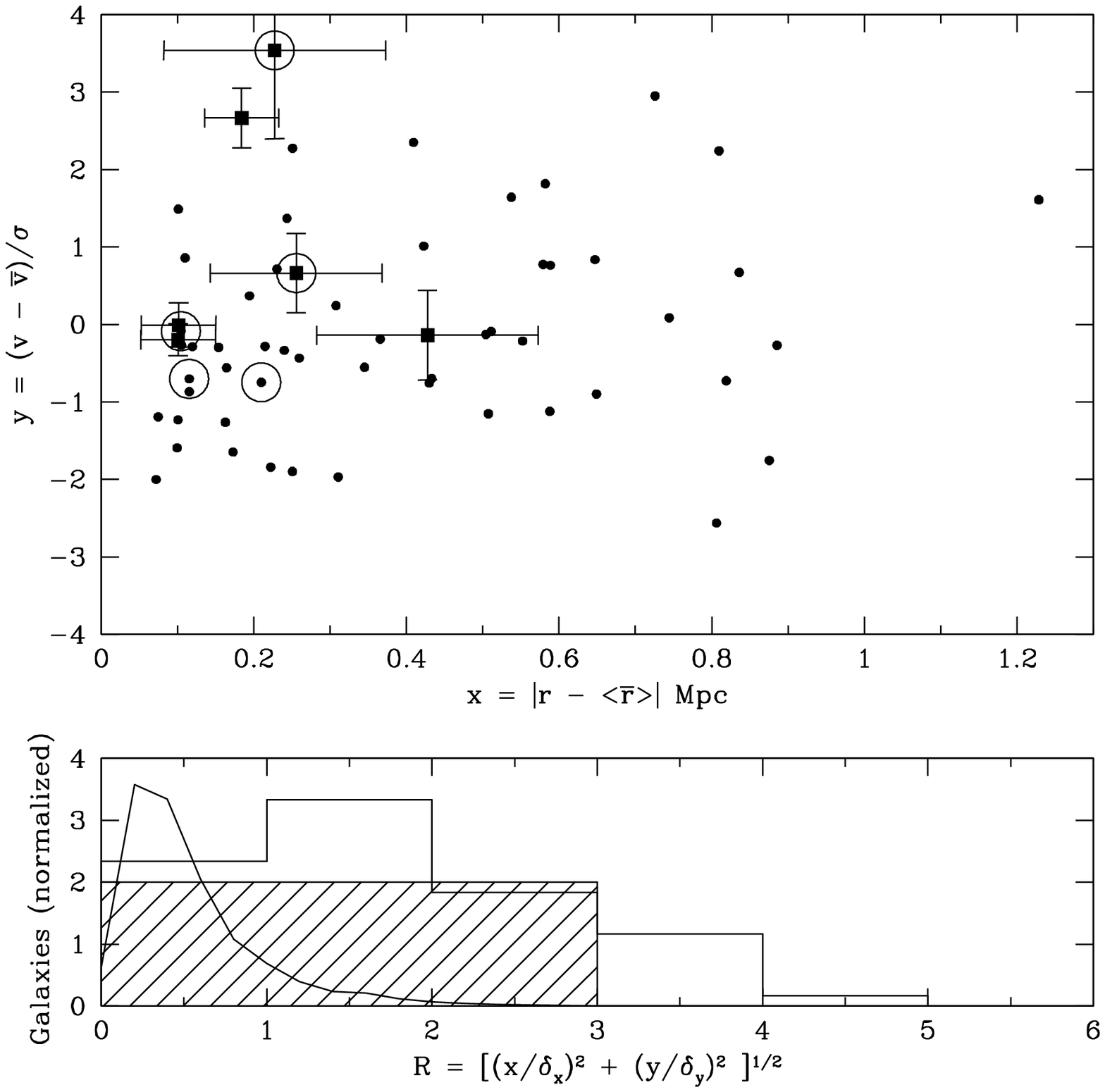}
\caption{
\emph{(Top)} Projected spatial and kinematic offsets of the brightest
group galaxy (BGG; filled squares) and all other group members
(filled circles) from the group centroid for five of our groups.
The lens galaxies are marked with open circles.  (The lens galaxy
in BRI0952 is omitted because it does not have a measured redshift.)
The velocity offset is normalized by the velocity dispersion of the
group to compensate for the differences among the group potentials.
The $y$ errorbars represent the 68\% confidence level based on adding
the errors for the lens galaxy velocity and the mean group velocity
in quadrature.  The $x$ error bars are the 68\% confidence level
based on a statistical bootstrap test (see text).
\emph{(Bottom)} Distribution of the phase-space offset $R$ for the
BGGs (shaded histogram) and all other group galaxies (open histogram,
normalized by the number of the brightest group galaxies). 
Statistical tests fail to distinguish between these two distributions.
The heavy line (scaled down by a factor of 400 to fit the y-axis)
shows the model $R$ distribution for a galaxy assumed
to lie at the bottom of the group potential (see text).  A KS-test
gives $7\times10^{-3}$ as the probability that the BGG and model
distributions are drawn from the same parent distribution, and a
t-test gives $7\times10^{-9}$ as the probability that the means of 
the BGG and model distributions are the same.  We conclude that the
BGGs generally do not occupy the center of the potential, and, that,
within the large uncertainties, their phase space distribution is
consistent with that of the other group members.
\label{rvsv}}
\end{figure}

The top panel of Figure \ref{rvsv} shows the projected spatial
$x$ and kinematic $y$ offsets from the group centroid for the
BGGs (filled squares) and for all other member galaxies (filled
circles) in our sample groups.  The $y$ errorbars ($\epsilon_y$)
represent the $1\sigma$ uncertainties based on adding the galaxy
velocity errors and the group mean velocity errors in quadrature.
The $x$ errorbars ($\epsilon_x$) represent the $1\sigma$
uncertainties based on the adding the centroid errors and the
individual galaxy position errors in quadrature (the centroid
errors dominate).  To estimate the statistical errors of the
centroid for a group of $N_{\rm grp}$ members, we carry out a
statistical bootstrap analysis where we draw 500 random samples
of $N_{\rm grp}$ galaxies without replacement from the B1422
group (the one with the most members), and adopt the variance of
the centroid position as its error.  For B1422, we use the smallest
of the errors calculated for the other groups.

We define the quantity $R^2 = (x/\delta_x)^2 + (y/\delta_y)^2$
as a measure of the phase-space distance of a galaxy from the
group centroid.  Here $\delta_x$ and $\delta_y$ are the rms
deviations in $x$ and $y$ for all galaxies plotted in the top
panel of Figure \ref{rvsv}.  A galaxy will have a large value
of $R$ if it has a large peculiar velocity and/or a position
that is far from the projected spatial centroid.  Conversely,
galaxies at rest in the center of the group potential will have
small $R$ values.  The bottom panel of Figure \ref{rvsv} shows
the distributions of $R$ values for the BGGs (shaded histogram)
and for all other group members (unshaded histogram).  We can
now use these distributions to answer two questions about the
BGGs.

{\em Are the BGGs distributed differently than the other group
galaxies?}  We compute the $R$ distributions for BGGs and for all
other group galaxies (see the bottom panel of Figure \ref{rvsv}),
and then compare them using three statistical tests: the KS-test
(to compare the overall distribution), the t-test (to compare the
means), and the F-test (to compare the variances).  All three fail
to distinguish between the two distributions.  In other words,
there are no significant differences between the orbits of BGGs
and the orbits of other group members, at least for these small
samples.  This result differs from observations of nearby groups
(see ZM98).  It is not clear from our present sample whether our
result indicates real evolution in the group galaxy population
between $z \sim 0.3$ and $z=0$, or is due simply to small number
statistics.  It will be interesting to return to this question
with the larger sample of lensing-selected groups that we are
obtaining.

{\em Are the BGGs consistent with the group centroids?}  We compare
the distribution of $R$ values for the BGGs with a model distribution
expected for a galaxy lying at the bottom of the group potential.
To incorporate measurement errors, we treat the model distribution
as a Gaussian in $x$ and $y$ with rms deviations of
$\epsilon_x/\delta_x$ and $\epsilon_y/\delta_y$, respectively, and
make 1000 random draws using the appropriate values of $\epsilon_x$
and $\epsilon_y$ for each group.  The bottom panel of Figure
\ref{rvsv} shows the observed and model distributions.  A KS-test
gives $6.6\times10^{-3}$ as the probability that the two samples
are drawn from the same distribution.  The probability that the
means of the two distributions are the same is $7\times10^{-9}$,
while the probability that the variances are the same is $10^{-5}$.
We conclude that the BGGs do not occupy the center of the group
potential.

This offset of the BGG from the kinematic and spatial centroid
of the group is not seen in nearby X-ray luminous groups.  While
our result could suggest group evolution from $z \sim 0.3$ to
now, another possibility is that we are not comparing apples to
apples.  For example, there are groups in our sample with velocity
dispersions lower than what is typical for dynamically-evolved
X-ray luminous groups nearby.  Among nearby groups,
lower-$\sigma_r$ systems tend to be dynamically younger and are
more likely to have an offset BGG (e.g., the Local Group).  This
may be true among our $z \sim 0.3$ groups as well.

Support for this latter interpretation comes from a closer look
at the BGGs in the three high-$\sigma_r$ groups in our sample:
B1422, PG1115, and MG0751.  The BGGs of MG0751 and PG1115 are
the galaxies closest to the projected spatial/kinematic centroid
in their respective groups, and, within the errors, are consistent
with being at the centroid.  In contrast, the BGG in B1422 lies
at a large projected distance from the projected spatial
centroid ($\sim\!0.22$  Mpc) and has a substantial peculiar
velocity (1270 \kms).  The BGG spatial and kinematic offset, together
with the clumpiness of B1422 (see \S 4.1), suggest that this system
is not yet relaxed.  Therefore, in two out of three cases, the
high-$\sigma_r$ groups in our sample are comparable to those nearby
groups with high velocity dispersions and centrally located BGGs.

We also compare the projected spatial centroid with the peak of
emission from the diffuse, luminous X-ray halos in PG1115 and B1422
(the two highest-$\sigma_r$ groups) seen by \citet{grantb1422-04}.
In PG1115, the projected spatial centroid and X-ray peak are
consistent within 2$\sigma$.  By contrast, in B1422 the peak of
the X-ray emission is substantially offset (more than 3$\sigma$)
from both the unweighted and luminosity-weighted group centroids. 

Our ability to address the questions of evolution raised in this
section will improve when we finish obtaining the larger sample of
lensing-selected groups.  Since lensing is sensitive to groups
spanning the redshift range $0.2 \lesssim z \lesssim 1$, it will
naturally provide the large redshift baseline needed to probe
evolution.  Working with a self-consistently selected sample of
groups will mitigate selection effects. 

\subsection{Group Contributions to Lens Potentials}

The understanding of environment-related biases and uncertainties in
lensing constraints on the masses and shapes of galaxy dark matter
halos, $H_0$, and substructure requires detailed lens modeling
(see KZ04).  That is beyond the scope of this paper and will be
treated separately.  For our purposes here, a simple way to
quantify environmental effects is to determine the dimensionless
convergence $\kappa$ and shear $\gamma$ that the group contributes
to the lens potential.  The convergence represents gravitational
focusing created by additional mass at the position of the lens
galaxy, while the shear represents tidal distortions created by
having an inhomogeneous distribution of matter near the lens
galaxy.  Convergence can never be constrained using lens models
alone because of the mass sheet degeneracy \citep{GFS,saha-degen}.
As a result, it is often omitted, which can lead to significant
biases in the model results (KZ04).  Shear cannot be constrained
in models of two-image lenses, which leads to enormous model
uncertainties.  While shear can be constrained in models of
four-image lenses, one of the puzzling results is that shear is
required in nearly all four-image lenses, and the required shear
strengths cannot easily be explained by traditional models of
large-scale structure \citep{keeton-97,dalal}.

Models of four-image lenses lead to the rule of thumb that a shear
of $\gamma \sim 0.1$ is common for groups, and $\gamma \sim 0.3$
for clusters.  Generally, we expect $\kappa \geq \gamma$ because
of the way convergence and shear add when there is more than one
perturber (see the Appendix).  Many lensing conclusions scale as
$(1-\kappa)$ to some power (see KZ04 and the Appendix), which can
help us estimate the biases.  For example, if a lens has convergence
$\kappa \sim 0.1$, then lens models that omit the convergence will
overestimate $H_0$ by $\sim$10\%.  Hence, we consider convergences
and shears larger than $\sim$0.1 to be quite important, and values
down to $\sim$0.05 worth consideration.

In this section, we quantify the effect of the group environments
in our sample by determining the convergence and shear due to the
group surrounding each lens galaxy.  Because it is not clear how
the mass is divided between the individual group members and a
common group dark matter halo, we consider two extreme cases that
bound the range of possibilities.  In the ``group halo limit,''
we assume that all of the mass is associated with a common group
halo, and we estimate the amount of mass from the velocity
dispersion.  In the ``group galaxies limit,'' we assume that all
of the mass is bound to the individual member galaxies, and we
calibrate the individual shears and convergences with recent weak
lensing observations \citep{sheldon}.  Comparing results from the
two limits can indicate how much our conclusions depend on how
the mass is distributed within the group.

\subsubsection{Group Halo Limit}

The group halo limit is appropriate for considering relaxed,
dynamically evolved groups in which the individual dark matter
halos of member galaxies may have been stripped by interactions,
so that the dominant component is a common dark matter halo.  In
this case, the luminous galaxies just trace the underlying mass
distribution, and the velocity dispersion of the galaxies is a
measure of the mass of the group as a whole.  If we model the
group halo as a singular isothermal sphere (SIS), then we can
take the velocity dispersion $\sigma_r$ of a group and its
centroid position $(b,\phi)$ in polar coordinates centered on the
lens galaxy, and compute the convergence $\kappa_{\rm grp}$ and
shear $\gamma_{\rm grp}$ using eq.~(\ref{SIS}) in the Appendix.
(Note that $\kappa_{\rm grp} = \gamma_{\rm grp}$ in the SIS
approximation.)  We then use the uncertainties in the velocity
dispersion and centroid position to determine the uncertainties
in the convergence and shear.  These derived uncertainties are
highly non-Gaussian, so we compute them using Monte Carlo
simulations.

The results are presented in
Table \ref{group.shear}.\footnote{As noted in the table caption,
we assume $z_s=2$ for the unknown source redshift in B2114, but
the particular value has little effect on our results.}
{\em In three of the six groups (MG0751, PG1115, and B1422), the
convergence and shear are significant.}  In these systems the
group environment needs to be accounted for in lens models, and
our measurements of the group properties make that feasible.  In
the other three groups (BRI0952, MG1654, and B2114), it appears
from our current data that the convergence and shear are small.

\begin{deluxetable*}{lccrrc}
\tablecolumns{6}
\tabletypesize{\footnotesize}
\tablecaption{Convergence and Shear in the Group Halo Limit\label{group.shear}}
\tablehead{ \colhead{Lens} & \colhead{$b$} & \colhead{$\sigma_r$} & \colhead{$\kappa_{\rm grp}=\gamma_{\rm grp}$} & \colhead{$\theta_\gamma$} & \colhead{Number of}\\ 
& \colhead{[$\arcsec$]} & \colhead{[\kms]} & & \colhead{[deg]} & \colhead{Images}}
\startdata
PG1115                 & 23 & 440 & {\bf 0.089} $_{-0.046}^{+0.065}$ & $  6\pm37$ & 4   \\
B1422                  & 43 & 470 & {\bf 0.058} $_{-0.024}^{+0.037}$ & $-65\pm20$ & 4   \\
MG0751                 & 23 & 320 & {\bf 0.049} $_{-0.033}^{+0.071}$ & $-27\pm36$ & 4+R \\
MG1654                 & 65 & 200 &      0.007  $_{-0.005}^{+0.011}$ & $-35\pm30$ & R   \\
BRI0952                & 61 & 170 &      0.005  $_{-0.004}^{+0.013}$ & $-81\pm42$ & 2   \\
B2114\tablenotemark{a} & 49 & 110 &      0.003  $_{-0.002}^{+0.014}$ & $-12\pm52$ & 2   \\
\enddata
\tablecomments{The groups are sorted by decreasing values of the
convergence and shear.  The magnitudes of the convergence and shear
are equal ($\kappa_{\rm grp} = \gamma_{\rm grp}$) in the SIS
approximation.  Values larger than 0.05 are marked in boldface.  The
angle $\theta_\gamma$ defines the direction of the shear (measured
North through East).  The convergence and shear errorbars are $1\sigma$
uncertainties derived from the uncertaintines in the group centroid
position and velocity dispersion (from Table \ref{groupproptable}).
Column 6 lists the image configuration for each lens: 2-image, 4-image,
or ring (R).}
\tablenotetext{a}{We assume a source redshift $z_s=2$ for B2114, but the
particular value has little effect on our results.  In particular,
assuming $z_s=3$ leads to the same numerical values for the
$\kappa_{\rm grp}$, $\gamma_{\rm grp}$, and $\theta_\gamma$.}
\end{deluxetable*}

The case of B1422 illustrates how our identifications of additional
group members affect estimates of the convergence and shear.
Based on six members, \citet{kundicb1422-97} estimated that the
group velocity dispersion is $\sigma_r = 550$ \kms, and that the
centroid lies a projected distance of $b \sim 14\arcsec$ from
the lens galaxy.  That led them to estimate a large ``observed''
shear of $\gamma_{\rm obs} = 0.23$, a value consistent
with the shear $\gamma_{\rm mod} \sim 0.2$ required by lens models
\citep{hoggb1422,keeton-97,dobler}.  Now with 16 members, we find
a smaller velocity dispersion $\sigma_r = 470$ \kms\ and a larger
offset $b = 43 \arcsec$, which reduce the nominal observed
shear to a more moderate value of $\gamma_{\rm obs} = 0.058$.
Accounting for the measurement uncertainties (which has not been
done before), we find allowed ranges of
$0.034 \le \gamma_{\rm obs} \le 0.095$ at $1\sigma$, and
$0.020 \le \gamma_{\rm obs} \le 0.170$ at $2\sigma$.  It is clear 
that omitting the group members can bias the estimates of how the 
environment affects the lens potential. However, even with an extensive 
catalog (16 members), uncertainties are important when comparing 
observations of the group with inferences from lens models.  
The large difference between the nominal observed shear 
$\gamma_{\rm obs} = 0.058$  and the shear required by models 
$\gamma_{\rm mod} \sim 0.2$ might not be significant at more 
than $2\sigma$. To test consistency between
the observations and models, it is best to make
new models with the environment constrained by our observations
(including the uncertainties, and possible clumpiness within the group; see \S 4.1).

\subsubsection{Group Galaxies Limit}

In the group galaxies limit, we suppose that all the mass in the
group is bound to the individual member galaxies.  This is probably
a better approximation for dynamically younger groups, which may
still be in the process of collapse and whose galaxies still retain
their halos.

To calibrate the shear and convergence from each galaxy, we turn to
observations that are perfectly suited to our needs: weak lensing.
The advent of large surveys, such as the Sloan Digital Sky Survey,
has made it possible to measure shear as a function of distance
from an average galaxy, on scales from 20 $h^{-1}$ kpc to
7 $h^{-1}$ Mpc \citep[e.g.,][]{sheldon}.\footnote{\citet{sheldon}
quote comoving distances, but we have converted to angular
diameter distances as those are more natural for our analysis.}
The projected offsets of galaxies in our sample range over
$\sim\!20$--700 $h^{-1}$ kpc, so we are working in precisely
the regime studied by \citet{sheldon}.  One possible concern with
this calibration is that the Sheldon analysis provides limited
information about how shear and convergence scale with luminosity
(or mass).  \citet{sheldon} measured the shear profile in three
luminosity bins, but most of our galaxies fall into just one of
the bins ($-22 < M_i - 5 \log h < -17$). Thus, we expect that
our present analysis characterizes the average properties of the
group galaxies well, but it will be worthwhile to redo the analysis
when future weak lensing data provide finer luminosity resolution. 

The quantity that is measured in weak lensing studies is shear,
but the observed shear profile is consistent with a power law
which means that there is a simple relation between the shear
and convergence.  The Appendix gives this relation, and also
provides more details about how we use the weak lensing data to
calibrate our analysis.

Once we have computed the convergence and shear from each galaxy,
we combine them in the manner of eqs.~(\ref{sum1})--(\ref{sum2})
in the Appendix.  The net convergence and shear for each group
are presented in Table \ref{galaxies.shear}.  We also list
statistical uncertainties derived (using Monte Carlo simulations)
from the errorbars on the weak lensing parameters quoted by
\citet{sheldon}.  These are generally small and are probably
less important than systematic effects due to incompleteness
(see below).  As in the group halo limit, in the galaxies limit
we find that MG0751, PG1115, and B1422 all have significant group
contributions to the lens potential.  The group contributions are
fairly small in MG1654, BRI0952, and B2114.  Comparing
Tables \ref{group.shear} and \ref{galaxies.shear} leads to a
crucial point: our conclusions about which groups significantly
affect the lens potentials do not depend on assumptions about how
the mass is distributed within the groups.  While the detailed
lensing implications do depend on the mass distribution --- which
is actually useful (see \S 4.4.3) --- it is reassuring to see that
our qualitative conclusions about which groups are important for
lensing are robust.

\begin{deluxetable*}{lrrrrrr}
\tablecolumns{4}
\tabletypesize{\footnotesize}
\tablecaption{Net Convergence and Shear in the Group Galaxies Limit\label{galaxies.shear}}
\tablehead{
  & \multicolumn{3}{c}{[Sheldon]} & \multicolumn{3}{c}{[$\sigma=100$ \kms]} \\
  \colhead{Lens} &
  \colhead{$\kappa_{\rm tot}$} & \colhead{$\gamma_{\rm tot}$} & \colhead{$\theta_\gamma$} &
  \colhead{$\kappa_{\rm tot}$} & \colhead{$\gamma_{\rm tot}$} & \colhead{$\theta_\gamma$} \\
& & & \colhead{[deg]} & & & \colhead{[deg]}
}
\startdata
PG1115                 & {\bf 0.066} $_{-0.006}^{+0.012}$ &      0.028  $_{-0.003}^{+0.009}$ & $ 64.2\pm3.2$           &      0.040  &      0.031  & $ 66$ \\
B1422                  & {\bf 0.092} $_{-0.009}^{+0.019}$ &      0.010  $_{-0.004}^{+0.012}$ & $-35.6_{-13.5}^{+25.4}$ & {\bf 0.064} &      0.019  & $-34$ \\
MG0751                 & {\bf 0.115} $_{-0.026}^{+0.040}$ & {\bf 0.091} $_{-0.034}^{+0.057}$ & $-82.9\pm1.8$           &      0.040  &      0.029  & $-88$ \\
MG1654                 &      0.027  $_{-0.004}^{+0.006}$ &      0.006  $_{-0.003}^{+0.005}$ & $ 21.4\pm13.2$          &      0.013  &      0.004  & $ 43$ \\
BRI0952                &      0.013  $_{-0.002}^{+0.002}$ &      0.007  $_{-0.001}^{+0.001}$ & $-64.6\pm8.3$           &      0.006  &      0.003  & $-67$ \\
B2114\tablenotemark{a} &      0.016  $_{-0.002}^{+0.004}$ &      0.012  $_{-0.002}^{+0.003}$ & $-15.9\pm3.3$           &      0.008  &      0.007  & $-16$ \\
\enddata
\tablecomments{The groups are sorted as in Table \ref{group.shear}.
Values larger than 0.05 are again marked in boldface.  Columns 2--4
list the net convergence and shear when we calibrate the individual
galaxies using the weak lensing observations by \citet{sheldon}.
The errorbars represent $1\sigma$ statistical uncertainties derived
from the weak lensing errorbars quoted by Sheldon et al.; even more
important may be systematic effects due to spectroscopic incompleteness
(see text).  For a simple comparison, Columns 5--7 list the results
when we treat each galaxy as an isothermal sphere with velocity
dispersion $\sigma = 100$ \kms.}
\tablenotetext{a}{Again, we assume a source redshift $z_s=2$ for B2114,
but the particular value has little effect on our results.}
\end{deluxetable*}

In carrying out this analysis we do not distinguish between
different morphological galaxy types.  Given their higher
mass-to-light ratios, elliptical (red) galaxies would produce
more shear and convergence than spiral (blue) galaxies of the
same luminosity.  To test the effects of morphology on our results
we consider the extreme assumption that all group galaxies are
red and use the \citet{sheldon} shear profiles for red galaxies.
In almost all cases the total convergence and shear due to the
group increase by $\sim$70\%.  Thus, if most group members are
red galaxies then our current analysis may actually underestimate
the convergence and shear by as much as 70\%.

As a simple sanity check, in Table \ref{galaxies.shear} we also
compute the net convergence and shear assuming that each galaxy
can be treated as a singular isothermal sphere (SIS) with velocity
dispersion $\sigma = 100$ \kms.  We expect that many of the galaxies
have larger velocity dispersions, and given that convergence and
shear scale as $\sigma^2$, this simple case should provide a
conservative lower bound on the net convergence from the group.
(The shear is more complicated, as discussed below.)  Comparing
the results assuming $\sigma=100$ \kms\ with those based on the
weak lensing calibration confirms this expectation, and generally
suggests that the weak lensing calibration is reasonable.  We
should note that the convergence from the simple SIS model may
not be a strict lower bound if halos are significantly truncated,
but in practice most of the convergence (and shear) arise from
galaxies close enough to the lens that truncation would have little
effect.  For example, if all halos had cut-off radii at 300 kpc,
the SIS model convergence would drop from $\kappa=0.040$ to 0.036
in PG1115, and from 0.064 to 0.057 in B1422.  Even with an extreme
cut-off at 100 kpc, the convergences would still be 0.027 for
PG1115 and 0.051 for B1422.  (In all cases, the shears are basically
unchanged.)  We expect that the conservative assumption of a small
velocity dispersion more than compensates for the omission of a
cut-off radius, so that the SIS model results are indeed lower
bounds.

It is again interesting to consider how our efforts to increase
the group membership have affected conclusions about the
convergence and shear.  The most instructive case is PG1115.
With our catalog of 13 members, we find a net convergence
$\kappa = 0.066_{-0.006}^{+0.012}$ and a net shear
$\gamma = 0.028_{-0.003}^{+0.009}$ at position angle
$\theta_\gamma = 64\pm3$.  If \citet{kundicpg1115} had done
the same analysis with their catalog of four members, they would
have found
$\kappa = 0.027_{-0.005}^{+0.008}$,
$\gamma = 0.014_{-0.003}^{+0.006}$, and
$\theta_\gamma = 69\pm5$.  (\citealt{tonrypg1115} would have
obtained similar results.)  In other words, the previous catalogs
missed at least half of the sources of shear and convergence.  The
problem was that they focused on the region within $\sim\!30\arcsec$
of the lens, but (in the galaxies approach) the galaxies have
extended dark matter halos, requiring the inclusion of group
members out to the full virial radius of the group in order
to fully characterize environmental effects in lens models.

With this thought in mind, we must consider how spectroscopic
incompleteness may affect the results in Table \ref{galaxies.shear}.
We cannot account for galaxies that we missed, but we can ask how
our results would have differed had we omitted a few of the galaxies
that we did actually include.  The effects of incompleteness on the
net convergence are simple: since convergences from different
galaxies combine in a simple scalar sum (see eq.~\ref{sum1} in the
Appendix), omitting galaxies causes us to underestimate the net
convergence.  Turning this around, we can say that our estimates
of $\kappa_{\rm tot}$ are strict {\em lower bounds} on the
convergence from the group.  To be more quantitative, in analyzing
subsamples of our group catalogs we find that the net convergence
scales roughly with the number of group members; so if the true
number of members is, say, 50\% larger than what we have observed
then we expect the true convergence to be $\sim$50\% larger than
our estimate.  For understanding biases in lens models, it is very
valuable to have a lower bound on the total convergence, because
that can be turned into lower bounds on the biases.

The shear is more complicated, because multiple contributions
sum as tensors rather than scalars (see eqs.~\ref{sum2c} and
\ref{sum2s}); adding more contributions can either increase or
decrease the net shear, and modify the position angle.  The
effects depend on the spatial distribution of member galaxies.
In PG1115 the most important galaxies lie roughly in a line
on the sky, which means that the direction of the net shear is
robust against incompleteness, while the amplitude of the shear
scales roughly with the number of group members.  In contrast,
in B1422 the galaxies are distributed more broadly, so
incompleteness may change the shear direction by tens of
degrees and the shear amplitude by tens of percent.  (Of
course, the shear from the galaxies in B1422 is small, so
even large fractional uncertainties are not so important.)
Finally, in MG0751 the environmental effects are dominated
by the galaxy G1 that is massive and close to the lens.

The bottom line is that incompleteness does not significantly
affect our qualitative conclusions.  Since our estimate of the
net convergence is a lower bound, we know that our conclusion
that groups are important on MG0751, PG1115, and B1422 is robust.
While it is possible that our conclusion that the groups are not so
important in MG1654, BRI0952, and B2114 could change if we
measure more galaxies and find that $\kappa$ rises, the observed correspondence between th group velocity dispersion and richness (see \S 4.2) suggests that adding more members would not affect $\kappa$ significantly.
Incompleteness issues will need to be considered when making
detailed quantitative comparisons between our environment
observations and lens models.

\subsubsection{Discussion}

It is important to understand the similarities and differences
between the results from the ``group halo'' and ``galaxies'' limits.
We have already noted that both limits lead to the same
conclusions about which groups are important for lensing.  All
three high-$\sigma_r$ groups (MG0751, PG1115, and B1422) produce
a significant convergence in both approaches, so these groups
cannot be ignored in lens models.  At the same time, two of the
low-$\sigma_r$ groups (BRI0952 and B2114) produce a small shear
and convergence in both approaches, suggesting that these groups
are not so important for lensing.  While this latter conclusion
may not seem exciting, it is actually quite valuable.  Two-image
lenses (including both BRI0952 and B2114) suffer from a strong
degeneracy between ellipticity and shear, if both quantities are
unknown.  That degeneracy can now be broken by ruling out models
with large shear.  The situation is less clear for MG1654, because
the group halo analysis implies negligible shear, while the galaxies
analysis yields a small but non-negligible shear $\gamma = 0.03$
(and that could be an underestimate).

When we turn to a more quantitative comparison of the two
approaches, we notice some significant differences.  The differences
suggest that it may be possible to distinguish between the group halo
and galaxies limits, and thus to learn about the distribution of
dark matter within the groups.  For example, in B1422 the group halo
analysis leads to a moderate convergence and shear, while the
galaxies analysis leads to a larger convergence but a negligible
shear.  Lens models require a large shear $\gamma \sim 0.2$ that
is marginally consistent with the group halo approach (given our
uncertainties) but grossly inconsistent with the galaxies
approach. In PG1115, as in B1422, lens models require large external shear $\gamma \sim 0.1$ \citep{impeypg1115}. The group halo limit result ($\gamma=0.089_{-0.046}^{+0.065}$) is within 1$\sigma$ of the model requirement while the galaxies limit gives a factor of three lower shear, inconsistent with predictions. The results in both B1422 and PG1115 --- the two highest velocity dispersion groups --- suggest that the mass is distributed in a common halo rather than being attached to the individual group members (as might be expected if high velocity dispersion groups are more dynamically evolved).  This hypothesis needs
to be examined more carefully with detailed lens models; rather
than just comparing the shear required by lens models with that
inferred from our observations, it is important to build models
that explicitly incorporate the environment (which may even consist
of multiple subgroup halos; see \S 4.1) in both the group halo and
galaxies limits and see whether either case can fit the lens
data.  Systems like B1422 and PG1115 may provide an exciting opportunity to
determine the distribution of dark matter in a distant group.

We also note that KZ04 used PG1115 as a
fiducial example with which to asses environment-related biases
in lens models.  \citet{dalal} suggested that PG1115 is a very
atypical lens environment, and that KZ04 therefore overestimated
environmental effects.  To the contrary, we find that PG1115 is
nothing if not typical: the group's kinematic properties and
shear and convergence are consistent with at least half of the
groups in our sample.

Another interesting system is MG0751.  Here, the galaxies
analysis is dominated by the G1 galaxy, lying just 6\arcsec\
from the lens.  Even so, \citet{Leharmg0751-97} showed that
lens models including only the lens galaxy, G1, and up to
three other nearby galaxies cannot fit the lens data.  It
will be interesting to use new lens models to test the
hypothesis that both G1 and the common group halo contain
significant mass, and to see whether we can constrain their
relative masses.  We must issue two warnings, however.  First, the projected offset of G1 from
the lens galaxy is just 20 $h^{-1}$ kpc, which lies at the
inner limit of the range studied by \citet{sheldon}; thus,
the reliability of the weak lensing calibration is not
clear.  Second, we show in \S 4.5.2 that MG0751 also has
a significant shear from a group along the line of sight,
which must be included in lens models along with the group
at the lens redshift.

In summary, our results above show that: (1) Group environments,
whether the mass lies with individual member halos (galaxies
limit) or in a common group halo (group halo limit), can contribute
significantly ($\kappa, \gamma \geq 0.05$) to the lens potential.
(2) If the members have halos (galaxies limit), they can have a
big effect, perhaps even greater than that of a common group halo.
(3) In the galaxies limit, correcting for incompleteness is only
going to boost the convergence (but will move the shears in either
direction).  (4) In the highest velocity dispersion groups, the shears produced in the group halo limit are more consistent with the observationally required values, suggesting that we might be able to discriminate between the models of the mass distribution.

Finally, it is remarkable that the three lenses with significant
environmental effects include both quad lenses (PG1115 and B1422)
and the one quad/ring (MG0751), whereas the double/ring (MG1654)
and the two double lenses (BRI0952 and B2114) all have small
convergence and shear.  While this result is limited by small
number statistics, it may suggest a correlation between image
configuration and environment.  Conventional wisdom
\citep[e.g.,][]{rusingroup} holds that shear does not
significantly affect the relative numbers of quads and doubles.
However, KZ04 argue that treating environment properly (including
terms beyond a simple shear) does change the quad/double ratio.
Our new results provide empirical evidence that there is a
connection.

\subsection{Lensing Effects of Line-of-Sight Structures}

Since lensing is a projected phenomenon, we must consider whether
structures projected along the line of sight significantly affect
strong lens systems.  Different theoretical approaches to studying
the effects of interlopers on strong lensing in a $\Lambda$CDM
universe have yielded contradictory results
\citep[e.g.,][]{barkana-96,keeton-97,premadi,wambs}, so an
empirical approach is clearly necessary.  To date, there are only
three lenses with confirmed line-of-sight groups
\citep[B0712+472, B1608+656 and MG 1131+0456;][]{fl,T&Kmg1131-00,fast2005} 
and several
other candidates \citep{faure,morgan}.  Here we present the first
systematic survey for structures along the line-of-sight to strong
lens systems.  The large redshift baseline and wide field of view
of our spectroscopic survey make it ideally suited to address this
issue.

\subsubsection{The Zone of Influence}

Our first task is to estimate the ``zone of influence'' for a
perturber along the line of sight to each lens in our sample.
Although it has not been done before, the calculation is
straightforward using the formalism presented in the Appendix.
Briefly, if we assume that a perturber can be modeled as an
isothermal sphere with some given velocity dispersion, then we
can use eqs.~(\ref{keff}), (\ref{geff}), and (\ref{SIS}) to
compute the effective convergence and shear ($\keff$ and $\geff$)
as a function of the impact parameter $b$ of the perturber relative
to the lens, and the redshifts of the lens galaxy and perturber.
(In the SIS approximation, $\keff=\geff$ for a single perturber.)

Figure \ref{sigma} shows contours of $\keff=\geff$ in the plane of
$b$ and $\Delta z = z_{\rm pert} - z_{\rm lens}$, for perturbers
with velocity dispersions of 100, 300, or 500 \kms.  (Because we do
not see rich clusters along the lines of sight to these lenses, this
range of $\sigma_r$ should span the observed range of structures.)
The grayscale is explained in the figure caption.  We consider the
zone of influence to be the region in which $\keff,\geff \geq 0.05$,
i.e., the unshaded (white) region in the figure.  The shape of this
region is not sensitive to the fact that an SIS halo has an infinite
extent; a cut-off halo radius of 300 kpc does not change the
zone of influence.  Also interesting are regions in which $\keff$
and $\geff$ go negative (shaded black), which represents a breakdown
of our formalism.  This happens only when the intrinsic convergence
of the perturber is $\kappa > 0.5$, which means that the offset
between the lens and perturber is small enough that the lens actually
lies within the Einstein radius of the perturber (see the Appendix).
In this case, the ``perturber'' is no longer just a perturbation
because its caustics interact with those of the main lens galaxy,
and we would observe a strong lensing effect from the second mass
as well.  This breakdown does not affect our conclusions because we
do not actually see any structures lying within this region; besides,
any objects that lie so close to the line of sight to the lens would
presumably be known from previous observations.

\begin{figure}
\centering
{\bf For figure see file f6\_sm.jpg}
\caption{\small
Contours of $\keff=\geff$ produced by a perturber with a velocity
dispersion of 100, 300 and 500 \kms, located at a redshift
$0 < z_{\rm pert} < 1$ and having an impact parameter $b$ with
respect to the lens, computed for seven of the lenses in our
sample.  (We exclude PMN2004 because its lens redshift is unknown.)
The horizontal axis represents the redshift difference between
the perturber and lens galaxy.  The vertical axis represents
the projected distance of the perturber from the lens, out to
$2\arcmin$.  Contours are drawn at $\keff = \geff = 0.001, 0.01,
0.05, 0.1$ and $0.5$, although not all of them are clearly visible
in all panels. The grayscale is as follows:
$\keff,\geff\geq0.05$ (white);
$0.05>\keff,\geff\geq0.01$ (light gray);
$0.01>\keff,\geff\geq0.001$ (medium gray)
and $\keff,\geff < 0.001$ (dark gray).
Important regions are $0.5 > \keff,\geff > 0.05$, i.e., the
areas in white.  At small impact parameter,
the perturbation approximation breaks down as we enter a regime
in which the perturber itself causes strong lensing (see text);
in our calculations, this leads to negative values of $\keff$
and $\geff$ (shaded in black).  Notice the strong dependence
on $\sigma$: the zone of influence scales as $\sigma^2$.
Massive perturbers can have large effects even when they lie
far from the lens.  Another striking feature is the very wide
redshift baseline in front and behind the lens over which the
perturber can cause a significant effect.
\label{sigma}}
\end{figure}

As the velocity dispersion of the perturber increases, the zone
of its influence grows dramatically $\propto \sigma_r^2$.  Consequently,
a more massive perturber can produce a large shear and convergence
even when offset from the lens; a perturber with
$\sigma_r \sim 500$ \kms\ (i.e., a rich group) can be offset by as
much as $1\arcmin$ and still produce $\gamma \sim 0.05$.  Another
striking feature of Figure \ref{sigma} is the very wide redshift
baseline in front of and behind the lens over which a perturber can
cause significant effects.  This result shows that it is crucial to
catalog not just mass structures in the immediate vicinity of the
lens galaxy, but also elsewhere along the line of sight, in order
to model the lenses accurately.   

\subsubsection{Prominent Interloping Structures}

To quantify line-of-sight effects for the lenses in our sample,
we consider the most prominent structures (those likely to be
groups and clusters) identified from our redshift catalog.  This
approach is analogous to the group halo limit for groups around lens
galaxies (\S 4.4.1), because it accounts for the dark matter in
massive bound structures.  To identify potentially important
line-of-sight structures, we show the redshift histograms again
in Figure \ref{los}, now shading only those galaxies that lie
within $1\arcmin$ of the lens.  We have over-plotted the curve
of the normalized shear strength $\keff/\kappa$ = $\geff/\gamma$
(eq.~\ref{normshear}) to give an indication of how the convergence
and shear vary with redshift.  From the discussion in \S 4.5.1,
we expect structures with large velocity dispersions, small
projected offsets from the lens, and/or small redshift offsets
from the lens galaxy to contribute most to the lens potential.
To be conservative, we select only those peaks in Figure \ref{los}
that: (1) have at least four members; (2) lie within $\Delta z$
such that $\keff/\kappa$ and $\geff/\gamma$ are $\gtrsim 0.5$;
and (3) have at least one member projected within $1\arcmin$ of
the lens. For every peak, we set pessimistic $3\sigma$ velocity
limits and use bi-weight estimators of location and scale to
calculate the mean velocity $\upsilon_i$ and the line-of-sight
velocity dispersion $\sigma_{r,i}$.  Based on the membership, we
then calculate the projected spatial centroid of the structure,
and its offset from the lens galaxy $b_i$ and position angle
$\phi_i$.  Finally, we can use eqs.~(\ref{keff}), (\ref{geff}),
and (\ref{SIS}) to determine the effective shear and convergence. 

\begin{figure*}
\includegraphics[height=7in,angle=270]{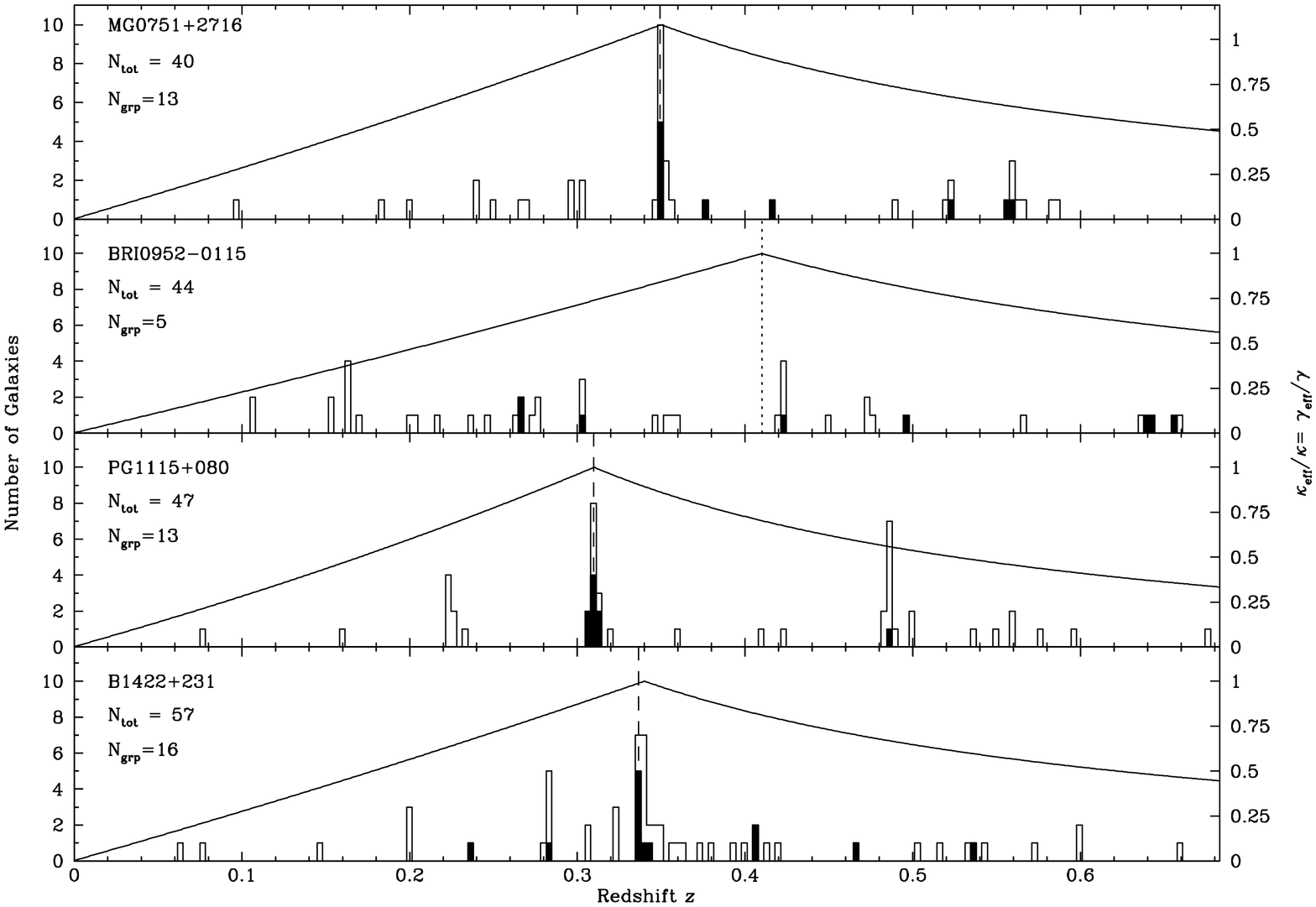}
\caption{
Same as the left panels in Figure~\ref{velocity}, except that
here we have shaded only those galaxies within 1\arcmin\ of the
lens.  We have also overplotted the normalized shear strength
and normalized convergence ($\keff/\kappa = \geff/\gamma$; see
text) to guide the eye regarding general behavior of convergence
and shear as the perturbing structure is moved away from the
lens in redshift.
\label{los}}
\end{figure*}

\begin{figure*}
\includegraphics[height=7in,angle=270]{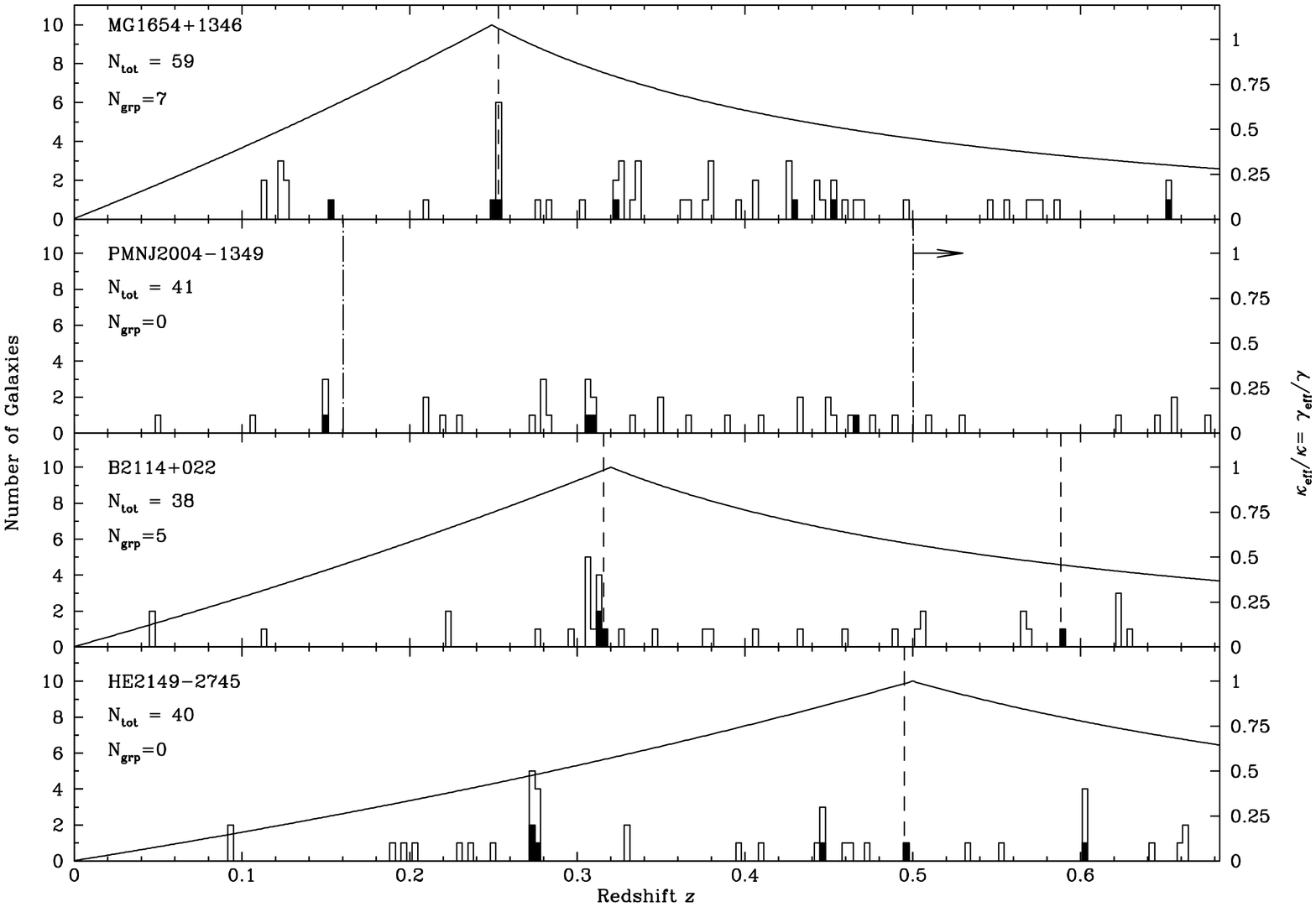}
\figurenum{\ref{los}}
\caption{continued.}
\end{figure*}

The results are presented in Table \ref{los.structures}.  We list
the individual effects of all prominent structures.  We choose to
include the groups at the lens redshifts (repeating results from
Table \ref{group.shear}), so that we can compare local versus
interloping structures.  We
then sum all structures (using eqs.~\ref{sum1}--\ref{sum2}) 
to obtain the total convergence and
shear for each lens (in the group halo limit).  We emphasize that
our results are conservative in the sense that we have not tallied
all the line-of sight objects that might affect the
lens models.  We do not include galaxies outside prominent peaks 
and plan to address this question in future work. We have not included 
all peaks in the velocity
histograms (Fig.~\ref{los}), because most are undersampled or
intrinsically poor, and thus have too few members for us to
interpret them as likely groups and to compute a meaningful
velocity dispersion. In addition, the velocity dispersions we
compute are probably underestimates not only because of the
narrow (conservative) choice of initial velocity ranges but
also because many of the velocity peaks are poorly sampled.
Finally, our spectroscopic target selection prioritizes galaxies
thought to lie at the lens redshift (see \S 3), and to some
extent that limits our ability to identify interloping structures.

\begin{deluxetable*}{lcccccrccr}
\tablecolumns{10}
\tablecaption{Convergence and Shear Due to Prominent Line-of-Sight Structures\label{los.structures}}
\tablehead{ \colhead{Lens} & \colhead{ID} &  \colhead{$z_{\rm pert}$} & \colhead{$b$} & \colhead{$\sigma_r$} & \colhead{$\keff=\geff$} & \colhead{$\theta_\gamma$} & \colhead{$\kappa_{\rm tot}$} & \colhead{$\gamma_{\rm tot}$} & \colhead{$\theta_{\gamma,{\rm tot}}$}\\
& & & \colhead{[$\arcsec$]} & \colhead{[\kms]} & & \colhead{[deg]} & & & \colhead{[deg]}
}
\startdata
MG0751  &   1   & 0.35 & 23 & 320 & {\bf 0.049} & $-27$ &             &             &         \\
MG0751  &   2   & 0.56 & 28 & 550 & {\bf 0.066} & $-56$ &             &             &         \\
MG0751  & total &      &      &     &             &         & {\bf 0.116} & {\bf 0.101} & $-44$ \\
\colrule
BRI0952 &   1   & 0.42 & 61 & 170 &      0.005  & $-82$ &             &             &         \\
BRI0952 & total &      &      &     &             &         &      0.005  &      0.005  & $-82$ \\
\colrule
PG1115  &   1   & 0.31 & 23 & 440 & {\bf 0.089} & $  7$ &             &             &         \\
PG1115  &   2   & 0.49 & 44 & 300 &      0.010  & $-74$ &             &             &         \\
PG1115  & total &      &      &     &             &         & {\bf 0.099} & {\bf 0.080} & $  5$ \\
\colrule
B1422   &   1   & 0.34 & 43 & 470 & {\bf 0.058} & $-65$ &             &             &         \\
B1422   &   2   & 0.28 & 80 & 400 &      0.020  & $-28$ &             &             &         \\
B1422   & total &      &      &     &             &         & {\bf 0.078} & {\bf 0.066} & $-57$ \\
\colrule
MG1654  &   1   & 0.25 & 65 & 200 &      0.007  & $-35$ &             &             &         \\
MG1654  & total &      &      &     &             &         &      0.007  &      0.007  & $-35$ \\
\colrule
HE2149  &   2   & 0.27 & 51 & 400 &      0.017  & $-62$ &             &             &         \\
HE2149  &   3   & 0.45 & 70 & 180 &      0.004  & $-80$ &             &             &         \\
HE2149  &   4   & 0.60 & 56 & 150 &      0.003  & $ 88$ &             &             &         \\
HE2149  & total &      &      &     &             &         &      0.024  &      0.022  & $-68$ \\
\colrule
B2114   &   1   & 0.31 & 49 & 110 &      0.003  & $-12$ &             &             &         \\
B2114   & total &      &      &     &             &         &      0.003  &      0.003  & $-12$ \\
\enddata
\tablecomments{``Prominent'' structures are defined as having at least
four members, at least one of which is projected within 1\arcmin\ of
the lens galaxy and displaced by $\Delta z$ such that
$\kappa_{\rm eff}/\kappa = \gamma_{\rm eff}/\gamma>0.5$.
Column 2 labels all the prominent structures along the line of sight
to each lens; ``0'' refers to a group at the lens redshift.  (HE2149
is the only lens with more than one interloping structure.)
Columns 3--7 refer to individual structures, while Columns 8--10 give
the final results after combining all the prominent structures along
the line of sight to each lens (including a group at the lens redshift,
if there is one).  Values larger than 0.05 are marked in boldface.}
\end{deluxetable*}

The discussion of incompleteness in \S 4.4.2 applies here as well,
with one minor change.  As before, we note that convergence is a
scalar and that the contribution from an overdensity is always
positive, and conclude that identifying additional prominent
structures would only {\em increase} the total convergence
$\kappa_{\rm tot}$.  The change is that an uncertainty in the
convergence zeropoint prevents us from declaring that we have
obtained a strict lower bound on the total convergence from the
line of sight. The zeropoint uncertainty arises from the effects of 
underdensities such as voids.
Strong lensing calculations conventionally assume that there are
a few density peaks superposed on top of a smooth background at
the mean density of the universe.  Voids can then be thought of
as regions where the density is negative (relative to the mean),
which contribute {\em negative convergence} to the lens potential
\citep[e.g.,][]{seljak-94}.  If there is a significant
negative convergence, it effectively changes the convergence
zeropoint: in summing the effects of overdense structures along
the line of sight, we should start not from zero but from the
appropriate negative convergence.

We cannot measure the convergence zeropoint directly because that requires 
knowledge of the local density of matter (dark and luminous) at every point 
along the line of sight, which is impossible to obtain even with complete 
redshift surveys. Nevertheless,
we can make a useful estimate based on a simple model.  
We create Monte Carlo simulations
of random lines of sight in a universe in which some of the mass
is contained in halos while the rest is in a smooth background at
a level below the mean density.  In these simulations we are able
to determine the total convergence $\kappa_{\rm tot}$ from density
fluctuations along the line of sight, as well as the contribution
$\kappa_{\rm peaks}$ from prominent structures ($\kappa_{\rm peaks}$ 
is analogous to what we compute above from our
data).  We can then interpret the difference,
$\kappa_{\rm zp} \equiv \kappa_{\rm tot} - \kappa_{\rm peaks}$,
as the convergence zeropoint.  (Details of the calculation, and
further discussion of the results, are given by \citealt{keetonLOS}.)

Considering many random lines of sight, we find that $\kappa_{\rm zp}$
has a mean of zero and a dispersion of $\lesssim 0.02$, for typical
lens and source redshifts and reasonable halo mass functions.  
Because the zeropoint uncertainty is small, we believe that our
$\kappa_{\rm tot}$ values above are probably lower limits on the total
convergence.  In other words, it is unlikely that voids contribute 
enough negative convergence to counter the positive $\kappa$ from prominent 
peaks in a given line of sight. 

The main result is that four of the eight lenses in our sample
have significant interloping structures.  MG0751, PG1115, and
B1422 each have one structure, and HE2149 has three.  At least
one of those structures, along the line of sight to MG0751, has
a significant contribution to the lens potential.  This
perturbing group lies at a redshift of $z=0.56$, which places
it between the lens galaxy and the source quasar, and has a
velocity dispersion $\sigma_r = 550$ \kms\ based on the six
members that we have identified.  With a centroid that lies at
a small projected offset of 28\arcsec\ from the lens, the group
contributes a convergence and shear
$\kappa_{\rm los} = \gamma_{\rm los} = 0.066$.  This is clearly
an important contribution to the lens potential --- in fact, it
is slightly stronger than our estimate of the contribution from
the group at the lens redshift (in the group halo limit).  It
represents one more piece in the interesting puzzle of fully
understanding lensing in the MG0751 system.

To our knowledge, our lens sample is not biased toward having
significant line-of-sight effects.  Our survey methods are, if
anything, somewhat biased {\em against} finding line-of-sight
structures (as discussed above).  Therefore, our discovery of
a significant structure in 1/8 lenses suggests that line-of-sight
effects are important in at least $\sim$10\% of all lenses.
That estimate needs to be confirmed with a larger sample, but
it does indicate that lensing effects from the line of sight
deserve further attention.

\section{Conclusions}

We have presented the  first results from our spectroscopic survey
of the environments of strong gravitational lenses.  We have used
multislit spectroscopy to measure the redshifts of 355 galaxies in
the fields of eight strong gravitational lenses with lens galaxies
at redshifts between 0.25 and 0.50.  After adding 16 redshifts from
the literature, we have analyzed a total sample of 371 galaxies with
redshifts.

The lens galaxy belongs to a poor group in six of the eight systems
in our sample.  We discover three new groups associated with the lens
galaxy of BRI0952 (five members), MG1654 (seven members), and B2114
(five members).  We more than double the number of members for another
three previously known groups around the lenses MG0751 (now 13 members),
PG1115 (13 members), and B1422 (16 members).  These six groups add to
the still small number of all poor groups identified at intermediate
redshifts.

We determine the kinematics of the six groups, including their mean
velocities, velocity dispersions, and projected spatial centroids.
For the newly discovered groups, we quantify these properties for
the first time.  For the other three groups, the increased membership
allows us to make more robust estimates of the kinematic properties
of the groups than previously possible.  The highest velocity
dispersions we measure (320 to 470 \kms\ for MG0751, PG1115, and
B1422) are consistent with those of nearby dynamically-evolved, X-ray
luminous groups (MZ98), while the lower velocity dispersions (110
to 200 \kms) are more typical of dynamically younger groups at low
redshift.  In the two cases where a diffuse X-ray component has
been measured \citep[PG1115 and B1422;][]{grantb1422-04}, the X-ray
temperatures and our velocity dispersions are consistent with the
local $\sigma_r$-$T_X$ relation (MZ98).

To understand the evolution of groups and their galaxies, it is
important to determine the relation of the brightest group galaxy
(BGG) to the group potential.  (In four of the six groups, MG0751,
BRI0952, PG1115, and B1422, the lens galaxy is not the BGG.)  We
find that the BGG generally lies off the center of the group
potential and occupies an orbit indistinguishable from the other
group members.  This result is surprising in comparison with
nearby, X-ray luminous groups, in which the BGG is always a giant
elliptical galaxy occupying the center of the potential, with an
orbit distinct from the other group members (ZM98).  However,
most of the effect we see comes from the three groups with lower
velocity dispersions.  In two (MG0751 and PG1115) of the three
highest velocity dispersion groups, the BGGs lie within the errors
of the group centroid, suggesting that at least these systems are
comparable to dynamically-evolved poor groups in the local universe.

We use our detailed observations of the groups to assess how
environments affect gravitational lens models.  A key ingredient
is an accurate determination of any offset between the lens galaxy
and the group centroid on the sky.  In MG0751, PG1115, B1422, and
MG1654, the lens galaxy is offset spatially from the group
centroid.  Obtaining a larger sample (which is underway) to
determine the full distribution of lens vs. group offsets will be
important for understanding how lens environments affect
statistical quantities such as the quad/double ratio and lensing
constraints on $\Omega_\Lambda$ (see KZ04).

To quantify environmental contributions to lens potentials in
more detail, we estimate the convergence (gravitational focusing)
and shear (tidal distortions) from each group.  We consider two
different models of the group mass distribution that bound the
extremes of dynamical states of groups.  The members of young
groups are likely to still have large dark matter halos and the
group mass may be dominated by the dark matter halos of the
individual member galaxies. In this approach, we calibrate the shear
and convergence from each galaxy based on observations of weak
lensing in the Sloan Digital Sky Survey \citep{sheldon}, and then
sum the contributions appropriately.  As the group evolves,
these halos may be truncated via interactions, so the mass will
be redistributed into a common group halo that may be the dominant
group mass component. In this approach, we approximate that halo as
an isothermal sphere and use the measured group centroid and
velocity dispersion to compute the convergence and shear.

At least three of the lenses in our sample (MG0751, PG1115, and
B1422) have convergences and shears large enough
($\kappa, \gamma \geq 0.05$) to indicate that the environment
plays a significant role in the lens potential.  For these systems,
our survey substantially improves the observational constraints
that will be needed to make detailed lens models that properly
include environmental effects.  Remarkably, the high shear and
convergence values occur in the quad lens systems, while the
environments of the double lenses are relatively weak.  This
result suggests that environment may affect the relative numbers
of quad and double lenses, a topic much debated in the literature
\citep[see][]{kingQD,csk96b,keeton-97,rusin,cohn,kz04}.  For the
other lenses, the conclusion that environment does not significantly
affect the lens potential is also valuable: constraining previously
unknown environmental terms to be near zero will still improve
lens models.

For the first time, we present a systematic assessment of whether
structures along the line of sight to lens systems are important
for lensing.  We show that interloping structures can in principle
affect lens models over a wide range of spatial and redshift offsets.
Our pencil-beam survey is ideally suited to identifying such
structures if they are present.  We find that at least four out of
eight lenses have prominent line-of-sight structures, i.e., groups
whose spatial and redshift offsets place them in the ``zone of
influence'' of the lens.  MG0751, PG1115, and B1422 each have one
substantial group along the line of sight, while HE2149 has three
groups at different redshifts.  Of these, the interloping group
in MG0751 has a significant effect on the lens potential.  Our
survey is actually biased against interloping groups (and is not
complete), so finding that at least one of eight lenses
($\sim\!10\%$) is affected by projected structures is intriguing
and worth further study.

\acknowledgments
We thank the staff of the Magellan Observatory for their tireless
efforts on behalf of this project.  We appreciate the advice and
spectral templates given by Marc Postman, as well as the helpful
comments provided by Chris Impey.  We thank the anonymous referee
for careful and constructive comments.
This work was supported by NSF grant \#AST-0206084 and NASA LTSA
award \#NAG5-11108.
IM acknowledges the support of the Martin F.\ McCarthy Scholarship
in Astrophysics awarded by the Vatican Observatory.

\appendix

\section{Formalism for Convergence and Shear}

\citet{keetonshear2003} presents a formalism for deriving a simple
analytic estimates of the effective convergence $\keff$ and shear
$\geff$ produced by a perturber somewhere along the line of sight
to a lens.  The same formalism can be used for all the situations
considered in the main text (see \S\S 4.4 and 4.5).  The only
assumption is that the convergence and shear are small, so that
we can work at first order in both quantities.  We will show below
that there is a context in which this assumption breaks down, but
with no effect on our conclusions.

If the perturber were at the same redshift as the lens galaxy, it
would produce a convergence and shear given by
\begin{eqnarray}
\label{keqn}
  \kappa &=& \frac{1}{2} \left(
    \frac{\partial^2 \varphi_{\rm pert}}{\partial x^2} +
    \frac{\partial^2 \varphi_{\rm pert}}{\partial y^2} \right)\,, \\
\label{gceqn}
  \gamma_c &=& \frac{1}{2} \left(
    \frac{\partial^2 \varphi_{\rm pert}}{\partial x^2} -
    \frac{\partial^2 \varphi_{\rm pert}}{\partial y^2} \right)\,, \\
\label{gseqn}
  \gamma_s &=& \frac{\partial^2 \varphi_{\rm pert}}{\partial x \, \partial y}\ , \\
  \gamma &=& \sqrt{\gamma_c^2+\gamma_s^2}\ , \\
  \theta_\gamma &=& \frac{1}{2}\tan^{-1}\left(\frac{\gamma_s}{\gamma_c}\right)\,,
\end{eqnarray}
where $\varphi_{\rm pert}$ is the lens potential of the perturber.
Here $\gamma$ is the shear strength, and $\theta_\gamma$ the shear
direction (which we measure North through East).  If the perturber
lies at a different redshift $z_{\rm pert}\neq z_l$, then the
convergence and shear are modified to the effective values
\citep{keetonshear2003}
\begin{eqnarray}
\label{keff}
  \keff &=& \frac{ (1-\beta) \left[ \kappa - \beta(\kappa^2-\gamma^2)
    \right] }{ (1-\beta\kappa)^2-(\beta\gamma)^2 }\ , \\
\label{geff}
  \geff &=& \frac{ (1-\beta) \gamma }
    { (1-\beta\kappa)^2-(\beta\gamma)^2 }\ ,
\end{eqnarray}
where 
\begin{eqnarray}
  \beta &=& \frac{D(z_1,z_2)}{D(0,z_2)}\,\frac{D(0,z_s)}{D(z_1,z_s)}\ , \\
  z_1 &=& \min(z_l,z_{\rm pert})\,, \\
  z_2 &=& \max(z_l,z_{\rm pert})\,,
\end{eqnarray}
where $D(z_1,z_2)$ is the angular diameter distance between redshifts
$z_1$ and $z_2$.  Note that when $z_{\rm pert} = z_l$, we have
$\beta=0$ and so we recover $\keff = \kappa$ and $\geff = \gamma$.
Moving the perturber away from the lens redshift increases $\beta$
and decreases $\keff$ and $\geff$.

In certain circumstances, the effective convergence and shear can
apparently go negative, which seems puzzling.  The sign flip occurs
only when the denominator in eqs.~(\ref{keff})--(\ref{geff}) goes
negative.  This can happen only when $\kappa$ and $\gamma$ are
sufficiently large --- in particular, only when the line of sight
passes through the {\em strong lensing} regime of the perturber.
(For an isothermal sphere [below], this corresponds to
$\kappa, \gamma \ge 0.5$.)  In this case, the lensing critical
curves of the perturber would merge with those of the main lens
galaxy, which would completely change the configuration of lensed
images.  In other words, our formalism breaks down when the
``perturbation'' is sufficiently strong, but if it were that strong
it would (presumably) be known already.

If $\kappa$ and $\gamma$ are small, then we can expand
eqs.~(\ref{keff})--(\ref{geff}) to first order write
\begin{equation}
\label{normshear}
  \frac{\keff}{\kappa} \approx \frac{\geff}{\gamma}
    \approx 1-\beta\,.
\end{equation}
We can think of $\keff/\kappa$ and $\geff/\gamma$ as the
``normalized'' convergence and shear --- the actual perturbation
strength, normalized by the value that would apply if the perturber
were at the same redshift as the lens galaxy.  We use this quantity
in the text as a simple way to characterize the redshift dependence
of the perturbation strength.

We now specify how to determine the net convergence and shear when
multiple perturbers are present.  This is relevant when we
assess the effects of a group around a lens by considering the
member galaxies (\S 4.4.2), and also when we consider the effects
of structures along the line of sight (\S 4.5.2).  We
see from eqs.~(\ref{keqn})--(\ref{gseqn}) that the quantities
$\kappa$, $\gamma_c$, and $\gamma_s$ are linear in the perturber
potential.  The shear is a combination of $\gamma_c$ and $\gamma_s$
that actually corresponds to a rank-2 traceless tensor, or a
headless vector (headless because it is invariant under rotation
by $180\arcdeg$).  (See \citealt{SEF} for more discussion.)  Thus,
if we know the effective convergence $\kappa_{{\rm eff},i}$ and
shear $\gamma_{{\rm eff},i}$, as well as the shear position angle
$\theta_{\gamma,i}$ (North through East), for a set of perturbers,
then the proper way to compute the net effects is as follows:
\begin{eqnarray}
\label{sum1}
  \kappa_{\rm tot} &=& \sum_i \kappa_{{\rm eff},i}\,. \\
\label{sum2c}
  \gamma_{c,{\rm tot}} &=& \sum_i \gamma_{{\rm eff},i} \cos 2\theta_{\gamma,i}\,, \\
\label{sum2s}
  \gamma_{s,{\rm tot}} &=& \sum_i \gamma_{{\rm eff},i} \sin 2\theta_{\gamma,i}\,, \\
  \gamma_{\rm tot} &=& \sqrt{\gamma_{c,{\rm tot}}^2+\gamma_{s,{\rm tot}}^2}\ ,\\
\label{sum2}
  \theta_{\gamma,{\rm tot}} &=& \frac{1}{2}\,\tan^{-1}({\gamma_{s,{\rm tot}}}/{\gamma_{c,{\rm tot}}})\,.
\end{eqnarray}
Note that $\kappa_{{\rm eff},i} \ge 0$ for any real perturber, so
the terms in the convergence sum all go in the same direction.  By
contrast, the cosine and sine factors mean that terms in the shear
sums may add or cancel.  Hence, incompleteness in our sample can
only cause us to underestimate $\kappa_{\rm tot}$, but it may cause
us to over- or underestimate $\gamma_{\rm tot}$.

It is worthwhile to recall how convergence and shear affect lens
models.  Convergence is largely responsible for systematic biases in
lens models (KZ04), through the mass sheet degeneracy
\citep{GFS,saha-degen}.  The biases can be thought of as simple
rescalings of model parameters, such as:
\begin{eqnarray}
\label{scaling}
  \beta &\propto& (1-\kappa_{\rm tot})\,,\\
  h     &\propto& (1-\kappa_{\rm tot})\,,\\
  \mu   &\propto& (1-\kappa_{\rm tot})^{-2}\,,
\end{eqnarray}
where $\beta$ is a mass parameter related to the lens velocity
dispersion, $h$ is the Hubble parameter, and $\mu$ is the image
magnification (see KZ04 and references therein).  Neglect of
convergence is the main source of biases in lensing results for
quad lenses.  Double lenses, by contrast, are so under-constrained
that poor knowledge of convergence and shear leads to lens models
that are just plain wrong.  In both cases, detailed observations of
lens environments are necessary to derive the constraints necessary
to make lens models reliable.

While the formalism presented so far is fully general, it is valuable
to discuss two particular perturber models.  First, if we can
approximate a perturber as an isothermal sphere then we can easily
relate measurable quantities to $\kappa$ and $\gamma$.  Specifically, 
for an isothermal sphere with velocity dispersion $\sigma$ and impact
parameter $b$, we have
\begin{equation}
\label{SIS}
  \kappa = \gamma = 1.44\times10^{-5} \left(\frac{1\arcsec}{b}\right)
    \left(\frac{\sigma}{\mbox{km/s}}\right)^2
    \frac{D(z_{\rm pert},z_s)}{D(0,z_s)}\ .
\end{equation}
We can measure the line-of-sight velocity dispersion $\sigma_r$ of
the group and the offset $b$ between the lens galaxy and group
centroid, use them to determine $\kappa$ and $\gamma$, and finally
fold in the redshift difference as above to determine $\keff$ and
$\geff$.

The second specific model we consider is a power law density profile
calibrated by weak lensing.  In a large and detailed analysis of
galaxy--galaxy weak lensing in the Sloan Digital Sky Survey,
\citet{sheldon} present the average shear as a function of radius
for a sample of 127,001 lens galaxies.  They find that the shear
profile is consistent with a simple power law
\begin{equation}
  \gamma(R) = \frac{A}{\Sigma_{\rm crit}}\ R^{-\alpha}\,,
\end{equation}
where $\Sigma_{\rm crit} = (c^2 D_{os})/(4\pi G D_{ol} D_{ls})$ is
the critical surface density for lensing, which carries all the
dependence on the lens and source redshifts through the angular
diameter distances $D_{ol}$, $D_{os}$, and $D_{ls}$ between the
observer, lens, and source.  \citeauthor{sheldon} tabulate the
values of $\alpha$ and $A$ for galaxies in three luminosity
bins.\footnote{\citet{sheldon} actually tabulate power law
parameters for the galaxy--mass correlation function rather than
the shear directly, but their formalism makes it straightforward
to convert back to $A$ and $\alpha$.  At any rate, those parameters
are more fundamental in terms of what they measure.  One additional
technical point is that Sheldon et al.\ quote lengths using comoving
distances, but we convert those to angular diameter distances in
our analysis.}  For any given galaxy that we observe, we convert
from our measured $I$-band apparent magnitude to absolute magnitude
in the SDSS $i$-band, using the color and $K$-corrections computed
with \citet{bruzual} spectral synthesis models.  (The luminosity
bins used by Sheldon et al.\ are wide enough that small systematic
uncertainties in the magnitudes do not shift galaxies between bins.)
We then look up the values of $A$ and $\alpha$ for each luminosity
from Table 2 of Sheldon et al.  We use the lens and source redshifts
to compute $\Sigma_{\rm crit}$ and hence $\gamma(R)$.  The last
thing we need is the convergence.  For a power law
$\gamma(R) \propto R^{-\alpha}$, there is a very simple relation
between shear and convergence:
\begin{equation}
  \kappa(R) = \frac{2-\alpha}{\alpha}\ \gamma(R)\,.
\end{equation}
Thus, it is straightforward to compute $\kappa$ and $\gamma$ for
each galaxy.  We can then factor in the redshift distance relative
to the main lens galaxy as described above.


\begin{thebibliography}{}

\bibitem[Allington-Smith et al.(1990)]{ldss} Allington-Smith, J.~R., Breare, J.~M., Ellis, R.~S., Parry, I.~R., \& Shaw, G.~D. 1990, \procspie, 1235, 691

\bibitem[Angonin-Willaime et al.(1993)]{angoninpg1115} Angonin-Willaime, M.-C., Hammer, F., \& Rigaut, F. 1993, Gravitational Lenses in the Universe, 85 

\bibitem[Augusto et al.(2001)]{augustob2114-01} Augusto, P., et al. 2001, \mnras, 326, 1007

\bibitem[Bar-Kana(1996)]{barkana-96} Bar-Kana, R. 1996, \apj, 468, 17 

\bibitem[Bar-Kana(1997)]{barkanapg1115} Bar-Kana, R. 1997, \apj, 489, 21 

\bibitem[Beers et al.(1990)]{statistical} Beers, T.~C., Flynn, K., \& Gebhardt, K. 1990, \aj, 100, 32

\bibitem[Bertin \& Arnouts(1996)]{bertin} Bertin, E., \& Arnouts, S. 1996, \aaps, 117, 393

\bibitem[Blandford et al.(2001)]{blandford} Blandford, R., Surpi, G., \& Kundi'c, T. 2001, in Gravitational Lensing: Recent Progress and Future Goals (ASP Conference Proceedings, vol. 237), ed. T. G. Brainerd \& C. S. Kochanek, p. 65

\bibitem[Browne et al.(1998)]{browne-98} Browne, I. W. A., Wilkinson, P. N., Patnaik, A. R., \& Wrobel, J. M. 1998, \mnras, 293, 257

\bibitem[Bruzual \& Charlot(1993)]{bruzual} Bruzual, G., \& Charlot, S. 1993, \apj, 405, 538

\bibitem[Burud et al.(2002)]{burudhe2149-02} Burud, I., et al. 2002, \aap, 383, 71

\bibitem[Carlberg et al.(2001)]{carlberg} Carlberg, R.~G., Yee, H.~K.~C., Morris, S.~L., Lin, H., Hall, P.~B., Patton, D.~R., Sawicki, M., \& Shepherd, C.~W. 2001, \apj, 552, 427 

\bibitem[Chae, Mao \& Augusto(2001)]{chaeb2114-01} Chae, K.-H., Mao, S., \& Augusto, P. 2001, \mnras, 326, 1015 

\bibitem[Chae(2003)]{chae-stats} Chae, K. 2003, \mnras, 346, 746

\bibitem[Chiba(2002)]{chiba} Chiba, M. 2002, \apj, 565, 17 

\bibitem[Cohn et al.(2001)]{cohn1933} Cohn, J. D., Kochanek, C. S., McLeod, B. A., \& Keeton, C. R. 2001, \apj, 554, 1216

\bibitem[Cohn \& Kochanek(2004)]{cohn} Cohn, J.~D., \& Kochanek, C.~S. 2004, \apj, 608, 25

\bibitem[Dalal \& Kochanek(2002)]{DK} Dalal, N., \& Kochanek, C. S. 2002, \apj, 572, 25

\bibitem[Dalal \& Watson(2004)]{dalal} Dalal, N. \& Watson, C. R. 2004, astro-ph/0409438

\bibitem[Danese et al.(1980)]{stddiv} Danese, L., de Zotti, G., \& di Tullio, G. 1980, \aap, 82, 322 

\bibitem[Dobler \& Keeton(2005)]{dobler} Dobler, G., \& Keeton, C. R. 2005, astro-ph/0502436

\bibitem[Fassnacht \& Lubin(2002)]{fl} Fassnacht, C.~D., \& Lubin, L.~M. 2002, \aj, 123, 627

\bibitem[Fassnacht et al.(2004)]{flb1608} Fassnacht, C. D., et al. 2004, Proceedings of IAU Symposium 225: Impact of Gravitational Lensing on Cosmology (also astro-ph/0409086)

\bibitem[[Fassnacht et al.(2005)]{fast2005} Fassnacht, C. D., et al., astro-ph/0510728

\bibitem[Faure et al.(2004)]{faure} Faure, C., Alloin, D., Kneib, J.~P., \& Courbin, F. 2004, \aap, 428, 741 

\bibitem[Ferreras et al.(2005)]{ferreras} Ferreras, I., Saha, P., \& Williams, L. L. R. 2005, astro-ph/0503168

\bibitem[Fischer et al.(1998)]{hst1411} Fischer, P., Schade, D., \& Barrientos, L.~F. 1998, \apjl, 503, L127

\bibitem[Fukugita et al.(1995)]{fukugita-color} Fukugita, M., Shimasaku, K., \& Ichikawa, T. 1995, PASP, 107, 945

\bibitem[Gerke et al.(2005)]{gerke} Gerke, B. F., et al. 2005, \apj, 625, 6

\bibitem[Gorenstein et al.(1988)]{GFS} Gorenstein, M.~V., Shapiro, I.~I., \& Falco, E.~E. 1988, \apj, 327, 693

\bibitem[Grant et al.(2004)]{grantb1422-04} Grant, C. E., Bautz, M. W., Chartas, G., \& Garmire, G. P. 2004, \apj, 610, 686

\bibitem[Hege et al.(1981)]{hegepg1115} Hege, E.~K., Hubbard, E.~N., Strittmatter, P.~A., \& Worden, S.~P. 1981, \apjl, 248, L1 

\bibitem[Henry \& Heasley(1986)]{hhpg1115} Henry, J.~P., \& Heasley, J.~N. 1986, \nat, 321, 139 

\bibitem[Hogg \& Blandford(1994)]{hoggb1422} Hogg, D.~W., \& Blandford, R.~D. 1994, \mnras, 268, 889

\bibitem[Holder \& Schechter(2003)]{holder} Holder, G.~P., \& Schechter, P.~L. 2003, \apj, 589, 688

\bibitem[Impey et al.(1996)]{impeyb1422-96} Impey, C. D., Foltz, C. B., Petry, C. E., Browne, I. W. A., \& Patnaik, A. R. 1996, \apj, 462, L53 

\bibitem[Impey et al.(1998)]{impeypg1115} Impey, C.~D., Falco, E.~E., Kochanek, C.~S., Leh{\' a}r, J., McLeod, B.~A., Rix, H.-W., Peng, C.~Y., \& Keeton, C.~R. 1998, \apj, 509, 551 

\bibitem[Keeton et al.(1997)]{keeton-97} Keeton, C.~R., Kochanek, C.~S., \& Seljak, U. 1997, \apj, 482, 604

\bibitem[Keeton \& Kochanek(1997)]{kk1115} Keeton, C. R., \& Kochanek, C. S. 1997, \apj, 487, 42

\bibitem[Keeton et al.(1998)]{keetonoptical-98} Keeton, C.~R., Kochanek, C.~S., \& Falco, E.~E. 1998, \apj, 509, 561

\bibitem[Keeton et al.(2000)]{kcz} Keeton, C.~R., Christlein, D., \& Zabludoff, A.~I. 2000, \apj, 545, 129

\bibitem[Keeton(2003)]{keetonshear2003} Keeton, C.~R. 2003, \apj, 584, 664

\bibitem[Keeton \& Zabludoff(2004)]{kz04} Keeton, C.~R., \& Zabludoff, A.~I. 2004, \apj, 612, 660

\bibitem[Keeton et al.(2005a)]{foldreln} Keeton, C.~R., Gaudi, B.~S., \& Petters, A.~O. 2005a, astro-ph/0503452

\bibitem[Keeton et al.(2005b)]{keetonLOS} Keeton, C. R., et al., 2005b, in preparation

\bibitem[King et al.(1996)]{kingQD} King, L.~J., Browne, I.~W.~A., \& Wilkinson, P.~N. 1996, IAU Symp.~173: Astrophysical Applications of Gravitational Lensing, 173, 191

\bibitem[King et al.(1999)]{kingb2114-99} King, L. J., Browne, I. W. A., Marlow, D. R., Patnaik, A. R., \& Wilkinson, P. N. 1999, \mnras, 307, 225

\bibitem[Kneib et al.(2000)]{rxj0911} Kneib, J., Cohen, J.~G., \& Hjorth, J. 2000, \apjl, 544, L35

\bibitem[Kochanek(1991)]{csk91} Kochanek, C.~S. 1991, \apj, 373, 354 

\bibitem[Kochanek(1995)]{csk95} Kochanek, C.~S.\ 1995, \apj, 445, 559 

\bibitem[Kochanek(1996a)]{csk96a} Kochanek, C.~S. 1996a, \apj, 466, 638

\bibitem[Kochanek(1996b)]{csk96b} Kochanek, C.~S. 1996b, \apj, 473, 595 

\bibitem[Kochanek et al.(2000)]{kochanekfundplane-00} Kochanek, C.~S., et al. 2000, \apj, 543, 131

\bibitem[Kochanek \& Schechter(2003)]{kochanek-schechter} Kochanek, C.~S. \& Schechter, P.~L. 2003, in Measuring and Modeling the Universes (Carnegie Observatories Astrophysics Series, vol. 2), ed. W.~L. Freedman (also astro-ph/0306040)

\bibitem[Kundi\'c et al.(1997a)]{kundicpg1115} Kundi\'c, T., Cohen, J.~G., Blandford, R.~D., \& Lubin, L.~M. 1997, \aj, 114, 507 

\bibitem[Kundi\'c et al.(1997b)]{kundicb1422-97} Kundi\'c, T., Hogg, D. W., Blandford, R. D., Cohen, J. G., Lubin, L. M., \& Larkin, J. E. 1997, \aj, 114, 2276

\bibitem[Kurtz \& Mink(1998)]{rvsao} Kurtz, M.~J., \& Mink, D.~J. 1998, \pasp, 110, 934

\bibitem[Langston et al.(1988)]{langstonmg1654-88} Langston, G. I., et al. 1988, \baas, 20, 1001

\bibitem[Langston et al.(1989)]{langstonmg1654-89} Langston, G. I., et al. 1989, \aj, 97, 1283

\bibitem[Leh\'ar et al.(1993)]{Leharmg0751-93} Leh\'ar, J., McMahon, R. G., Irwin, M. 1993, AAS, 183, 3307L

\bibitem[Leh\'ar et al.(1997)]{Leharmg0751-97} Leh\'ar, J., et al. 1997, \aj, 114, 48L

\bibitem[Linder(2004)]{linder} Linder, E.~V. 2004, \prd, 70, 043534

\bibitem[Lopez, Wucknitz \& Wisotzki(1998)]{lopezhe2149-98} Lopez, S., Wucknitz, O., \& Wisotzki, L. 1998, \aap, 339, L13

\bibitem[Mao \& Schneider(1998)]{mao} Mao, S., \& Schneider, P. 1998, \mnras, 295, 587 

\bibitem[McMahon \& Irwin(1992)]{mcmahonbri0952-92} McMahon, R. \& Irwin, M. 1992, Gemini, 36, 1M

\bibitem[Metcalf \& Madau(2001)]{MM} Metcalf, R. B., \& Madau, P.. 2001, \apj, 563, 9

\bibitem[Mitchell et al.(2005)]{mitchell} Mitchell, J.~L., Keeton, C.~R., Frieman, J.~A, \& Sheth, R.~K. 2005, \apj, 622, 81

\bibitem[Morgan et al.(2005)]{morgan} Morgan, N.~D., Kochanek, C.~S., Pevunova, O. \& Schechter, P.~L. 2005, AJ, 129, 2531

\bibitem[Mulchaey \& Zabludoff(1998)]{zabmulch2} Mulchaey, J.~S.~\& Zabludoff, A.~I. 1998, \apj, 496, 73

\bibitem[Omont et al.(1996)]{omontbri0952-96} Omont, A., McMahon, R.~G., Cox, P., Kreysa, E., Bergeron, J., Pajot, F., \& Storrie-Lombardi, L.~J. 1996, \aap, 315, 1 

\bibitem[Patnaik et al.(1992a)]{patnaik-92} Patnaik, A. R., Browne, I. W. A., Wilkinson, P. N., \& Wrobel, J. M. 1992a, \mnras, 254, 655

\bibitem[Patnaik et al.(1992)]{patnaikb1422-92} Patnaik, A. R., Browne, I. W. A., Walsh, D., Chaffee, F. H., \& Foltz, C. B. 1992, \mnras, 259, 1P 


\bibitem[Postman et al.(2002)]{templates} Postman, M., Lauer, T.~R., Oegerle, W., \& Donahue, M. 2002, \apj, 579, 93

\bibitem[Premadi \& Martel(2004)]{premadi} Premadi, P., \& Martel, H. 2004, \apj, 611, 1

\bibitem[Refsdal(1964)]{refsdal} Refsdal, S. 1964, \mnras, 128, 307

\bibitem[Rusin et al.(2001)]{rusingroup} Rusin, D., et al. 2001, \apj, 557, 594

\bibitem[Rusin \& Tegmark(2001)]{rusin} Rusin, D., \& Tegmark, M. 2001, \apj, 553, 709

\bibitem[Rusin et al.(2003)]{rusin-03} Rusin, D., et al. 2003, \apj, 587, 143

\bibitem[Rusin \& Kochanek(2005)]{rusin-05} Rusin, D., \& Kochanek, C. S. 2005, \apj, 623, 666

\bibitem[Saha(2000)]{saha-degen} Saha, P. 2000, \aj, 120, 1654

\bibitem[Schechter et al.(1997)]{schechterpg1115} Schechter, P.~L., et al. 1997, \apjl, 475, L85 

\bibitem[Schneider et al.(1992)Schneider, Ehlers, \& Falco]{SEF} Schneider, P., Ehlers, J., \& Falco, E. E. 1992, Gravitational Lenses (Berlin: Springer)

\bibitem[Seljak(1994)]{seljak-94} Seljak, U. 1994, \apj, 436, 509

\bibitem[Sheldon et al.(2005)]{sheldon} Sheldon, E., et al. 2004, AJ, 127, 2544

\bibitem[Soucail et al.(2001)]{mg2016+112} Soucail, G., Kneib, J.-P., Jaunsen, A.~O., Hjorth, J., Hattori, M., \& Yamada, T. 2001, \aap, 367, 741

\bibitem[Tonry(1998)]{tonrypg1115} Tonry, J.~L. 1998, \aj, 115, 1 

\bibitem[Tonry \& Kochanek(1999)]{T&Kmg0751-99} Tonry, J. L., \& Kochanek, C. S. 1999, \aj, 117, 2034

\bibitem[Tonry \& Kochanek(2000)]{T&Kmg1131-00} Tonry, J. L., \& Kochanek, C. S. 2000, \aj, 119, 1078

\bibitem[Treu \& Koopmans(2004)]{TK04} Treu, T., \& Koopmans, L.~V.~E. 2004, \apj, 611, 739

\bibitem[Turner(1990)]{turner-90} Turner, E.~L. 1990, \apjl, 365, L43

\bibitem[van Waerbeke \& Mellier(2003)]{cosshear} van Waerbeke, L., \& Mellier, Y. 2003, astro-ph/0305089

\bibitem[Wambsganss, Bode \& Ostriker(2004)]{wambs} Wambsganss, J., Bode, P., \& Ostriker, J. P., 2004, astro-ph/0405147

\bibitem[Williams et al.(2005)]{kurtis} Williams, K., Momcheva, I., Keeton, C. R., \& Zabludoff, A. I. 2005, in preparation

\bibitem[Wilman et al.(2005)]{wilman} Wilman, D.~J., et al. 2005, astro-ph/0501183

\bibitem[Winn, Hall \& Schechter(2003)]{winnpmn2004-03} Winn, J. N., Hall, P. B., \& Schechter, P. L. 2003, \apj, 597, 672

\bibitem[Winn et al.(2001)]{winnpmn2004-01} Winn, J. N., Hewitt, J. N., Patnaik, A. R., Schechter, P. L., Schommer, R. A., L{\'o}pez, S., Maza, J., \& Wachter, S. 2001, \apj, 121, 122

\bibitem[Wisotzki et al.(1996)]{wisotskihe2149-96} Wisotzki, L., K{\"o}hler, T., Lopez, S., \& Reimers, D. 1996, \aap, 315, L405

\bibitem[Weymann et al.(1980)]{weymannpg1115} Weymann, R.~J., et al. 1980, \nat, 285, 641 

\bibitem[Wilkinson et al.(1998)]{wilkinson-98} Wilkinson, P. N., Browne, I. W. A., Patnaik, A. R., Wrobel, J. M., \& Sorathia, B. 1998, \mnras, 300, 790

\bibitem[Yahil \& Vidal(1977)]{3sigmaclip} Yahil, A.~\& Vidal, N.~V. 1977, \apj, 214, 347

\bibitem[Young et al.(1981a)]{youngpg1115} Young, P., Deverill, R.~S., Gunn, J.~E., Westphal, J.~A., \& Kristian, J. 1981, \apj, 244, 723 

\bibitem[Young et al.(1981b)]{young0957} Young, P., Gunn, J.~E., Oke, J.~B., Westphal, J.~A., \& Kristian, J. 1981, \apj, 244, 736

\bibitem[Zabludoff \& Mulchaey(1998a)]{zabmulch1} Zabludoff, A.~I., \& Mulchaey, J.~S. 1998, \apj, 496, 39

\bibitem[Zabludoff \& Mulchaey(1998b)]{substructure} Zabludoff, A.~I., \& Mulchaey, J.~S.\ 1998, \apjl, 498, L5

\end{thebibliography}
\end{document}